\documentstyle[graphicx,aps]{revtex}
\newlength\twdthfull
\newlength\twdthhalf
\newlength\twdththird
\newlength\twdthtwothird
\newlength\twdthquart
\newcommand{\srt}{SrTiO$_3$}
\newcommand{\tc}{$T_{c}$}
\newcommand{\gdnb}{$\gamma_{b}/\nu_{b}$}
\newcommand{\gdns}{$\gamma_{s}/\nu_{s}$}
\newcommand{\gs}{$\gamma_{s}$}
\newcommand{\ns}{$\nu_{s}$}
\newcommand{\gb}{$\gamma_{b}$}
\newcommand{\nb}{$\nu_{b}$}
\newcommand{\Dq}[1]{$\Delta$q$_{\rm #1}$}
\newcommand{\ang}[1]{$\times$10$^{#1}$\AA$^{-1}$}

\newcommand{\qvec}{{\mathbf q}}
\newcommand{\qvecp}{{\mathbf q'}}
\newcommand{\qvecn}{{\mathbf q_0}}
\newcommand{\kvec}{{\mathbf k}}

\newcommand{\shirane}{floatzone grown}
\newcommand{\mue}{$\mu$m}
\def\bilder{}

\begin{document}
%
%
\preprint{\sf Physical Review B} 
\title{Influence of defects on the
  critical behaviour at the \boldmath{$105$}\,K structural phase
  transition of SrTiO$_3$: II. The sharp component}
\author{H.~H\"unnefeld, T.~Niem\"oller and J.~R.~Schneider}
\address{Hamburger Synchrotronstrahlungslabor HASYLAB at Deutsches
  Elektronen-\\
  Synchrotron DESY, Notkestr. 85, D-22603 Hamburg, Germany}
\author{U.~R\"utt}
\address{Argonne National Laboratory\\
  9700 S.~Cass Avenue, Argonne, IL 60439, USA} \author{S.~Rodewald,
  J.~Fleig}
\address{Max-Planck-Institut f\"ur Festk\"orperforschung\\
  Heisenbergstr. 1, D-70569 Stuttgart, Germany} 
\author{G.~Shirane}
\address{Department of Physics, Brookhaven National Laboratory\\
  Upton, New York 11973, USA}
\date{June 7, 2000}
\maketitle
\pacs{PACS: 68.35.Rh, 61.10.-i, 64.60.-i, 68.35.Dv}
\begin{abstract}
  The depth dependence of the crystallographic parameters mosaicity,
  lattice parameter variation and integrated reflectivity and of the
  critical scattering above the $105$\,K structural phase transition
  of \srt{} have been studied in five different single crystals by
  means of high resolution triple-crystal diffractometry using
  $100-120$\,keV synchrotron radiation.  Depth-dependent impedance
  measurements indicate that the presence of oxygen vacancies is not
  responsible for the two-length scale phenomenon.  It is found that
  the sharp component occurs only in surface near regions of highly
  perfect single crystals and is coupled to an exponential inrease of
  the crystallographic quantities. The second length scale is absent
  at a surface where the strain fields are able to relax by a
  macroscopic bending of the lattice planes.  The sharp component is
  also strongly suppressed in crystals of relatively large
  mosaicity.  The combination of long range strain fields in highly
  perfect samples and the vicinity of the surface seem to be
  necessary conditions for the observation of the sharp component.
  The critical exponents for the second length scale are in satisfying
  agreement with scaling laws if the intensity of the critical
  scattering is assumed to be proportional to the square of the
  Lorentzian susceptibility and not, as usual in the current
  convention, to a Lorentzian-squared susceptibility.  The critical
  exponents of the broad component are independent of the appearance
  of the sharp component.
\end{abstract}
\newpage
\section{Introduction}
According to modern theories of phase transitions based on the concept
of scaling, the properties of the fluctuations associated with
continuous phase transitions close to the critical temperature \tc{}
are determined by a single length scale which varies with temperature.
In contrast high resolution X-ray diffraction studies of structural
phase transitions in perovskites revealed the existence of a second
length scale in the critical fluctuations above \tc{} which had not
been observed in earlier neutron scattering experiments.  Later this
effect has also been found in magnetic systems.  The two length scale
problem was discussed by Cowley \cite{Cow96a}.  The effect was first
observed at the cubic-to-tetragonal second-order phase transition in
\srt{} at $T_c\approx 105$\,K and the temperature dependence of the
two components has been studied by McMorrow et al.~\cite{McM90} usind
$11$\,keV synchrotron radiation.  Shirane et al.~\cite{Shi93} showed
that this sharp component is different from the central peak
discovered by Riste et al.~\cite{Ris71} in a study of the soft phonon
mode.  It has been shown that the sharp component originates from a
volume element close to ($\approx 100$\,\mue) the surface of the
samples \cite{Thu94,Neu95b,Rue97a} and is related to the quality of
the respective surfaces \cite{Thu94,Wat96}.  However, the nature of
the defects responsible for the sharp component is not clear,
e.g.~Wang et al.~\cite{Wan98a} suggested the existence of edge
dislocations close to the surface as to be responsible for the
occurrence of the second length scale.

In this paper we present a systematic investigation of the critical
scattering in different samples, differing by the respective growth
technique, the amount of oxygen vacancies and the crystallographic
quality.  Using a beam of high energetic photons of $100-120$\,keV and
$\sim 10$\,\mue{} in height surface effects could be separated
from bulk behaviour.  The effect of the defects on the critical
behaviour in the bulk of the samples has already been discussed in
\cite{Hue00a}, hereafter referred to as paper~I.  In order to clarify
the nature of the sharp component and the influence of defects on the
occurrence of the second length scale the depth dependence of the
crystallographic parameters mosaicity, lattice parameter variation and
integrated reflectivity has been analysed and compared to the depth
dependence of the critical behaviour.

After the description of the experimental setup of the diffraction
experiments in section~\ref{sec:diff}, the characterisation of the
different samples is reported in detail in section~\ref{sec:sam}.  The
measurements of the critical behaviour require a detailed
determination of the critical temperatures [section~\ref{sec:tc}]. The
experimental results for the critical exponents are summarised in
section~\ref{sec:crit}.  Finally, the results are discussed in
section~\ref{sec:sum}.  As an appendix mathematical details with
regard to the resolution function are given in section~\ref{sec:app}.
\section{Experimental setup}
\label{sec:diff}
The diffraction experiments have been performed with triple crystal
diffractometers at the beamlines BW5 and PETRA2 at HASYLAB in Hamburg,
using high energetic photons ($E\ge100$\,keV).  The details of the
experimental stations are described in \cite{Bou98} and
\cite{HAS95,HAS96,Oes00}, respectively.  Due to the high photon energy
of the incident beam, the photons penetrate through the crystal,
i.e.~all measurements have been performed in Laue-geometry.  The
critical scattering has been observed at the ($511$)/2-superlattice
reflection position.  Annealed silicon single crystals reflecting from
($311$) lattice planes were used as monochromator and analyser.  The
instrumental resolution in the scattering plane has been determined at
the position of the superlattice reflection a few degrees below the
transition temperature, examples are shown in
figure~\ref{fig:resolution}.  It results to \Dq{x}=1\ang{-3}
(\Dq{x}=1.5-3\ang{-3}) at PETRA2 (BW5) in the longitudinal direction
[fig.~\ref{fig:resolution}a] and 1.6\ang{-4}$\le$\Dq{y}$\le$2\ang{-3}
in the transverse direction [fig.~\ref{fig:resolution}b], depending on
the mosaicity of the respective sample.  Perpendicular to the
scattering plane the resolution (HWHM) was of the order of
\Dq{z}=1\ang{-1}.  A detailed description of the deconvolution of the
experimental data is described in the appendix.  The crystallographic
perfection has been characterised at the ($511$) main reflection, with
monochromator and analyser crystals using ($624$) reflections of
perfect silicon.  The difference in lattice spacing between sample and
monochromator/analyser is $\sim 3.6$\%.  Due to this almost
dispersion-free setup the instrumental resolution is improved both in
the longitudinal direction (\Dq{x}=1.2\ang{-4}) and in the transverse
direction (\Dq{y}=1.0\ang{-5}).

In addition to the high {\bf q}-space resolution also high real space
resolution could be achieved with sufficient count ratio realising a
narrow cross-section of the incident beam by means of a micro-slit and
taking advantage of the high incident photon flux generated by the
respective insertion devices at the beamlines.  The spot
size was reduced to a minimum height of $10$\,\mue{} and a width of
$2$\,mm.  The surface of the sample was arranged parallel to the beam
profile, the experimental procedure for the alignment of the
micro-slit is given in \cite{Rue97a}.  By vertical translation of the
sample the scattering volume can be moved to well-defined positions in
the sample [see figure~\ref{fig:shirane}].  Due to the special
geometry of the samples (the ($511$) reciprocal lattice vector is
almost parallel to the investigated surface) it is possible to study
the depth dependence of crystallographic quantities like strain and
mosaicity with a spatial resolution of $\sim 10$\,\mue.  In the
following, the depth~$0$\,\mue{} corresponds to the situation, where
the complete beam just penetrates the sample. Consequently, at a depth
of $-10$\,\mue{} the beam is just passing by the sample.  These
positions are identified by accurately measuring the dependence of the
transmitted intensity as a function of the vertical position of the
beam in the sample.
\section{Samples}
\label{sec:sam}
The influence of defects on the critical behaviour of \srt{} has been
studied in various samples, differing in growth technique and heat
treatments, which has been reported in paper~I.  In this paper we
concentrate on the results for two highly perfect single crystals,
namely sample~I, which has been grown by the top-seeded method
\cite{Sha72} more than two decades ago, and sample~II, grown by the
flux-grown technique \cite{Sche76}.  Additionally, a series of
Verneuil-grown samples has been investigated, an overview can be found
in table~\ref{tab:samples}, details with regard to the sample
preparation are given in paper~I.

Searching for the origin of the sharp component, which occurs
essentially close to the surface of the samples~I \cite{Rue97a} and
II, first the depth dependence of strain, mosaicity and integrated
reflecting power of these samples has been measured.  As discussed in
paper~I, the existence of point defects, in this case oxygen
vacancies, changes the values of the critical exponents if the mean
distance of the defects is smaller than the correlation length of the
critical fluctuations.  Therefore the depth dependent concentration of
the two-fold negatively charged oxygen vacancies has been investigated
using impedance spectroscopy \cite{Mac87}:
As described in \cite{Fle96} microelectrodes of different diameters can 
be used to analyse the depth dependence of the conductivity.
Using gold microelectrodes ($15-220$\,\mue) the conductivity measured 
at $\sim 200^\circ$C could be shown to be depth-independent in the top 
$500$\,\mue of both samples I and II.
Oxygen vacancies considerably contribute to the overall conductivity
at these temperatures and the conductivity data can be transferred 
into oxygen vacancy concentration values \cite{Den95a}.
It can therefore be concluded that the vacancy concentration is 
depth-independent as well.
Consequently, the sharp component, which has been
observed in sample~I up to a depth of $100$\,\mue{} \cite{Rue97a}, is
not related to a gradient in the oxygen vacancy concentration.  This
is an important finding because the sharp component has only been
found in the crystallographically rather perfect samples with 
relatively high concentrations of oxygen vacancies and not in the 
less perfect Verneuil crystals with much lower oxygen vacancy 
concentrations.  
In the Verneuil-grown samples the vacancy concentration of the bulk
has been determined from conventional impedance measurements of the 
isolating as-grown and oxidised samples using the vacancy mobility 
\cite{Den95a} and by Hall-resistivity measurements for
the almost metallic reduced sample.  The results are summarised in
table~\ref{tab:samples}.

\begin{figure}[h]
  \centering \includegraphics[width=\linewidth]{\bilder
    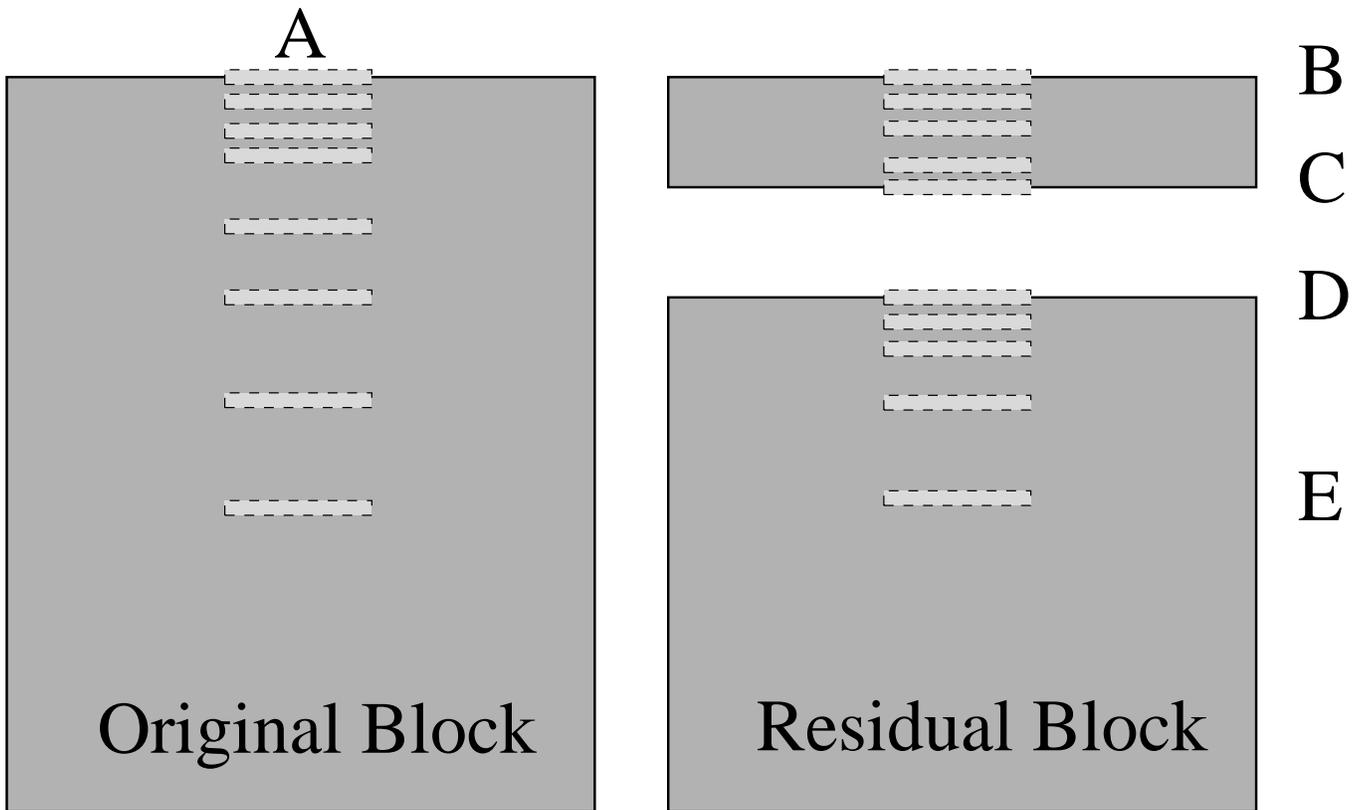}
    \caption{Schematic drawing of the \shirane{} sample. The left hand side shows the original sample investigated in [Rue97], the right hand side shows the two samples obtained after cutting a $560$\,\mue{} thick platelet from the top of the original sample. The capital letters A-E define the nomenclature for this paper. Region A corresponds to the surface of the original block. Region B and C correspond to the two surfaces of the platelet, region D denotes the surface of the residual block and region E labels the bulk of this block. The lighter rectangles (not to scale) indicate locations of the incident beam with respect to the sample surface. \label{fig:shirane}} 
\end{figure}

The main focus of this paper is on sample~I, where a direct
correlation between strain, lattice parameter variation and the
occurrence of the sharp component in the critical scattering was
observed in the surface near region up to a depth of $100$\,\mue, also
using triple crystal diffractometry at energies of $\sim 120$\,keV
\cite{Rue97a}.  Based on these results a $560$\,\mue{} thick slice has
been cut from the same sample.  Figure~\ref{fig:shirane} shows a
schematical drawing of the sample before and after cutting.  On the
left hand side the original sample is plotted.  The capital letter A
marks the region investigated in \cite{Rue97a}.  The rectangular spots
visualise the beam cross-sections on the sample, however, the
dimensions are not to scale.  The sample size is about $1$\,cm$^3$,
whereas the cross-section of the beam was $10$\,\mue{}$\times 2$\,mm.
The cut was performed parallel to the surface in a depth of about one
millimeter, using a diamond saw.  The residual plate has a thickness
of $\sim 560$\,\mue, the material loss due to sawing and careful
polishing of the two cut faces resulted to about $1$\,mm.  In
figure~\ref{fig:shirane} also the notation for the following
discussion is defined.  Region~B is the surface near region of the
upper face of the residual plate, which has not been polished or
changed in any way compared to the former measurements in region~A in
\cite{Rue97a}.  The region at the lower surface of the plate is
labeled~C.  This surface corresponds to a depth of $560$\,\mue{} with
respect to the original sample surface.  The region on top of the
residual block is labeled~D, originally this part of the crystal was
in a depth of $\sim 1.5$\,mm from the surface of the original block,
i.e.~it represented the bulk of the original sample where no sharp
component was observed.  The two surfaces C and D had been subject to
the identical polishing treatment.  Region~E represents the bulk of
the block.  The depth dependent impedance measurements described above
have been performed in region~A, at the original sample.

\begin{figure}[h]
  \centering \includegraphics[angle=-90,width=\linewidth]{\bilder
    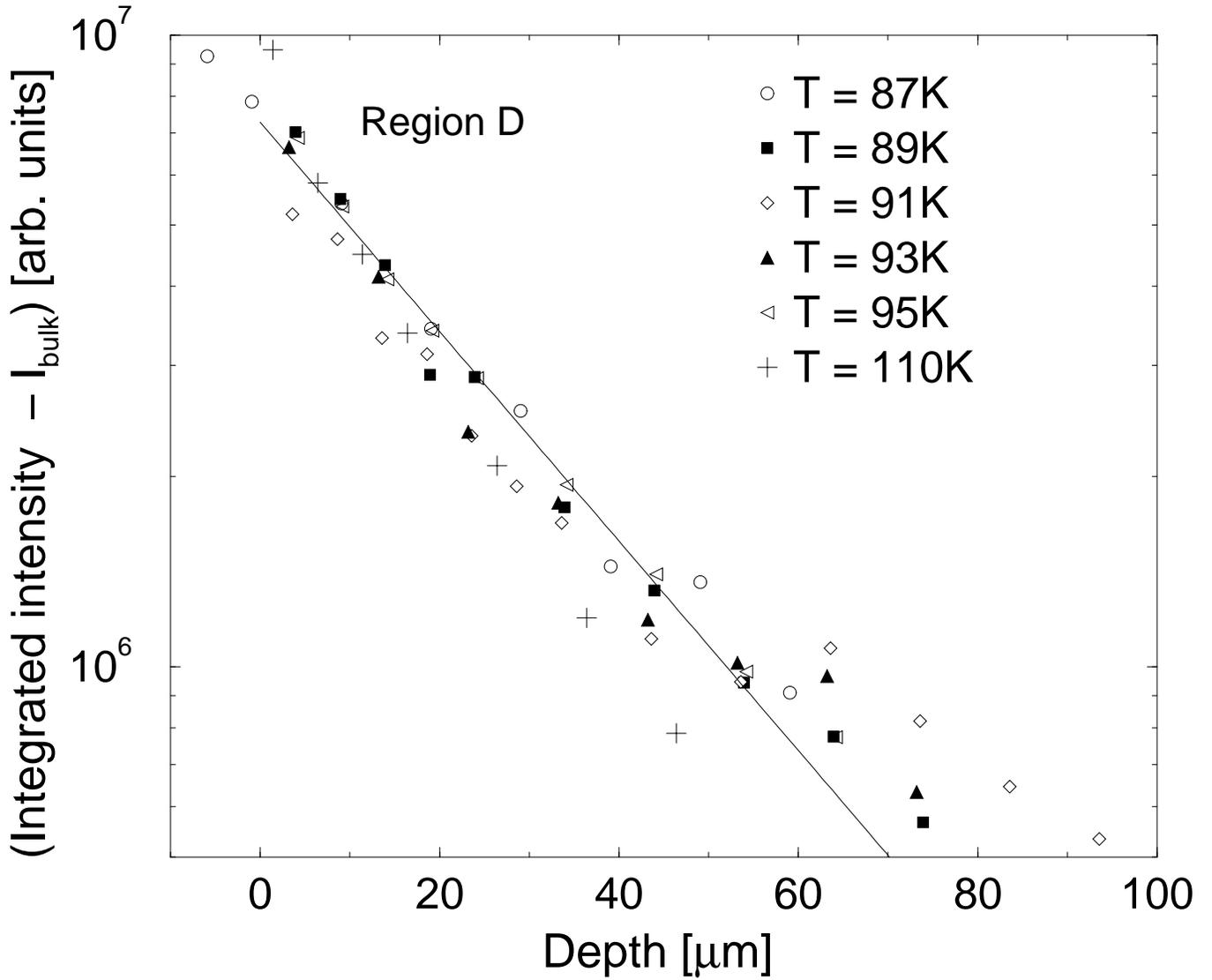}
    \caption{Depth dependence of the integrated intensity of the ($511$)-reflection in region~D for different temperatures around the critical temperature with respect to the corresponding bulk value $I_{bulk}$. The increase of the integrated intensity in the surface near region is well described by an exponential relation with a $1/e$-length of $\zeta=26(1)$\,\mue{}. \label{fig:int_D}}
\end{figure}

\begin{figure}[h]
  \centering \includegraphics[angle=-90,width=\linewidth]{\bilder
    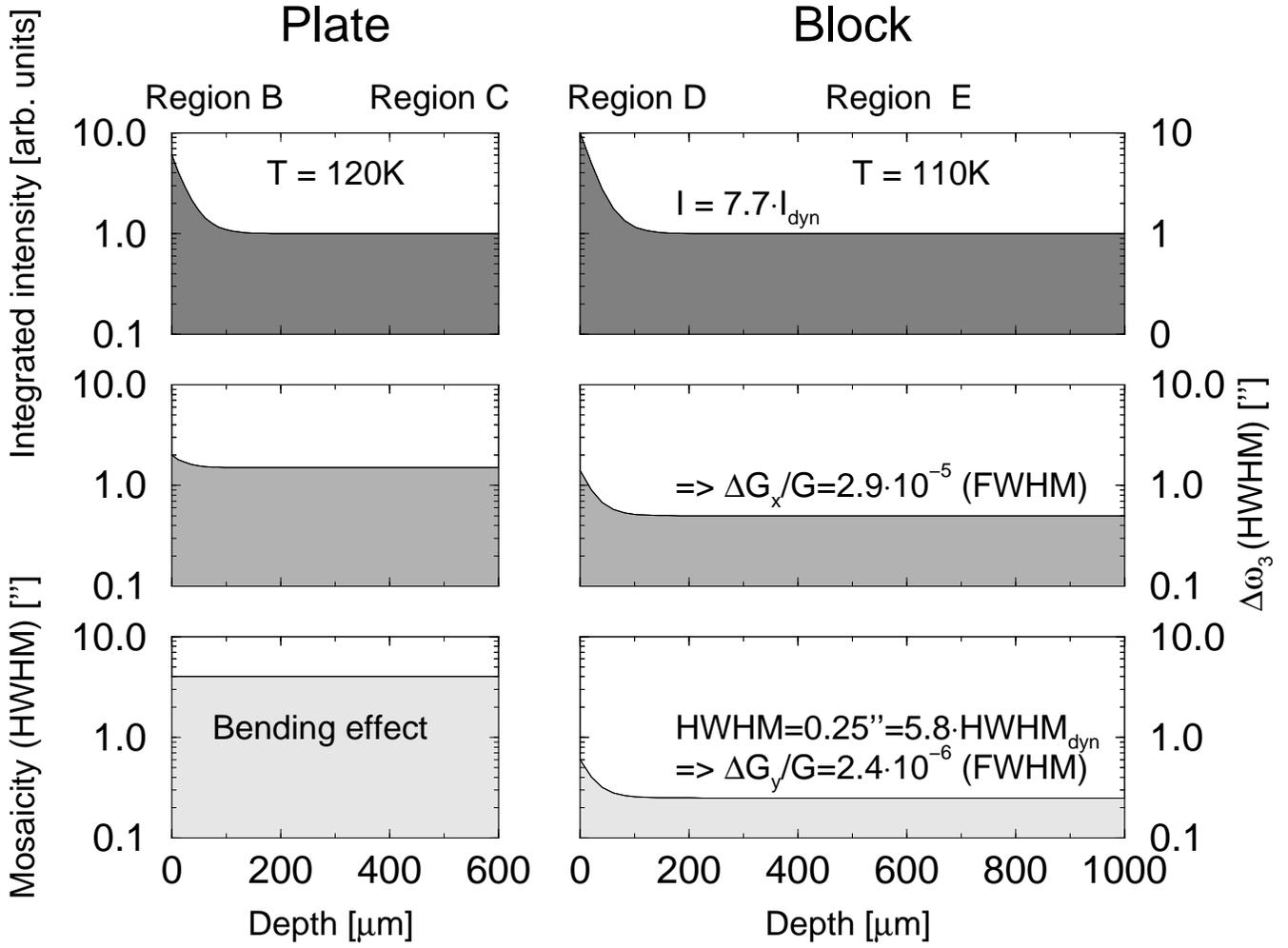}
    \caption{Schematic view of the depth dependence of integrated intensity and the widths (HWHM) of longitudinal and transverse scans at the ($511$) reflection position. The surface of the residual block (D) and the old surface of the plate (B) exhibit the same features, whereas the new surface of the plate (C) shows no effects in the crystallographic quantities. The intrinsic mosaicity of the plate (picture in the lower left corner) could not be measured due to the bending of the plate. The decay of the crystallographic parameters at the different surfaces is well described by exponential functions with the same $1/e$-length $\zeta\approx 25.5(15)$\,\mue. \label{fig:char}}
\end{figure}

The depth dependence of the integrated intensity, the mosaicity and
the variation of the lattice parameter has been measured at the
($511$) main reflection around the phase transition temperature of
$\sim 100$\,K for the different surfaces B--D and in the bulk~E of
sample~I.  In figure~\ref{fig:int_D} the gain in the integrated
intensity of the ($511$) reflection in region~D compared to the
intensity $I_{bulk}$ in the bulk (region~E) is plotted as a function
of the distance to the surface of the residual block.  The straight
line is an exponential function $(I-I_{bulk}) \propto \exp(-z/\zeta)$
fitted to the data.  No temperature dependence can be observed, the
mean $1/e$-length results to $\zeta =26(1)$\,\mue.  The increase of
intensity can be understood qualitatively on the basis of dynamical
diffraction theory \cite{Zac45,Bat64}.  The bulk mosaicity of the
sample is determined by transverse scans in reciprocal space, it
results to a FWHM of $\Delta\omega_2=0.5$''.  The width of the ($511$)
reflection of an absolutely perfect \srt-crystal at $100$\,keV is
$FWHM_{dyn}=0.086$'', i.e. the mosaicity is about a factor of $6$
larger than the dynamical width.  This broadening is not an effect of
the instrumental resolution but represents the intrinsic mosaic spread
of the sample, which is still much smaller than in the other samples,
especially the Verneuil grown samples with mosaic spreads of
$30-100$''.  The integrated reflectivity in the bulk expected for a
perfect crystal can be calculated for negligible absorption using
dynamical theory.  One obtains $I_{dyn}/I_0=3.2\times 10^{-7}$, where
$I_0$ is the transmitted intensity behind the sample in the anguar
range where no Bragg scattering occurs, i.e.~the effect of absorption
is eliminated.  In the bulk (region~E) the measured integrated
reflectivity $I/I_0$ was about $7.7$ times larger than the theoretical
value for a perfect crystal.  This is consistent with the increase in
the mosaic spread.  However, using the kinematical scattering theory,
the expected integrated reflectivity of the $12$\,mm thick sample
should be $I_{kin}/I_0=7.0\times 10^{-5}\approx 220 \cdot
I_{dyn}/I_0$.  Thus the diffraction mechanism in the bulk of this
sample is close to the expectations for a perfect crystal, which has
been shown before by means of $\gamma$-ray diffraction experiments
\cite{Sch86}.  Close to the surface the mosaic spread of the sample
increases [Fig.~\ref{fig:char}], and therefore the scattering process
has to be described more and more by kinematical scattering theory,
which explains the increase of the integrated reflectivity, shown in
figure~\ref{fig:int_D}.  The widths (HWHM) of the transverse scans
($\Delta\omega_2$), corresponding to the mosaicity, and the widths
(HWHM) of the longitudinal scans ($\Delta\omega_3$), corresponding to
the variation of the lattice parameter ($\Delta d/d =
\frac{1}{2}\cdot\cot\theta_B\cdot\Delta\omega_3$, $\theta_B$ is the
Bragg-angle) follow the same exponential depth dependence as the
integrated intensity shown in figure~\ref{fig:int_D}, the results are
plotted fig.~\ref{fig:char}.

\begin{figure}[h]
  \centering \includegraphics[angle=-90,width=\linewidth]{\bilder
    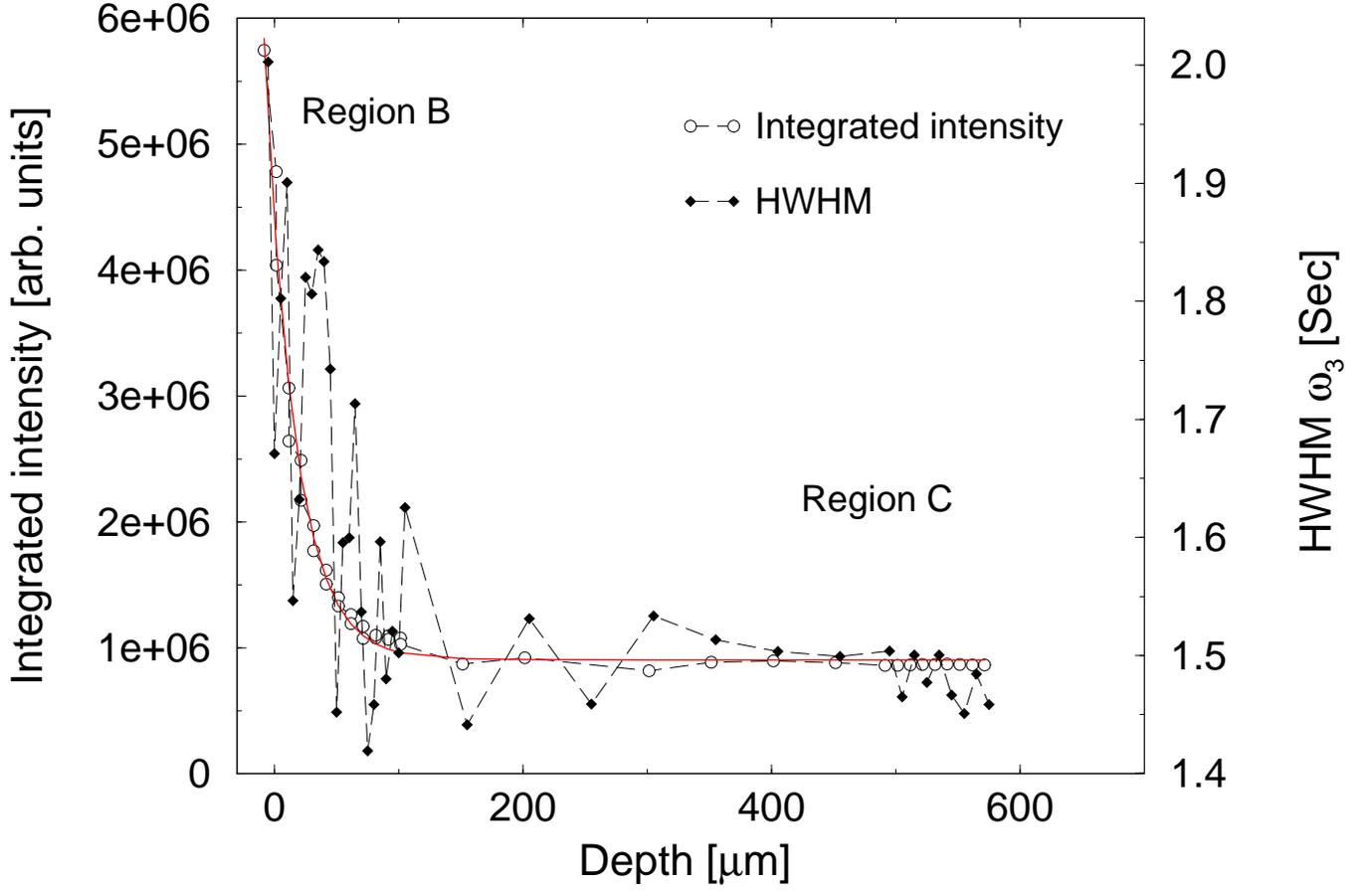}
    \caption{Depth dependence of the integrated intensity and the HWHM of the longitudinal scans at the ($511$)-reflection position in the plate. The width of the longitudinal scans is proportional to the lattice parameter variations: $\Delta d/d=\frac{1}{2}\cdot\cot\theta_B\cdot\Delta\omega_3$. The left hand side corresponds to region~B, an exponential increase ($(I-I_{bulk})\propto \exp(-z\zeta)$) of both quantities is clearly visible. The $1/e$-length results to $\zeta=25(1)$\,\mue. On the right hand side (region~C) no changes at all can be observed. \label{fig:int_BC}}
\end{figure}

The identical characterisation of the crystallographic properties was
carried out for the platelet (regions B and C).
Figure~\ref{fig:int_BC} shows the data of the integrated intensity in
the platelet and simultaneously the width (HWHM) of the longitudinal
scans, i.e.~the lattice parameter variation, at a temperature of
$120$\,K.  On the left side of the figure, which corresponds to
region~B, the behaviour is identical to that in region~D, shown in
figure~\ref{fig:int_D}.  Within the errorbars, the $1/e$-length
($\zeta=25(1)$\,\mue) is the same.  But, surprisingly, the other
surface of the plate (region~C) does not show any effect, neither for
the integrated intensity, nor for the widths of longitudinal and
transverse scans.  In fact, the integrated reflectivity is identical
to the bulk value in region~E, whereas the variation of the lattice
parameter, $\Delta d/d$ is slightly enhanced compared to the bulk
value.  These results are also summarised in figure~\ref{fig:char}.

\begin{figure}[h]
  \centering \includegraphics[width=\linewidth, bb= 50 415 522 780,
  clip]{\bilder 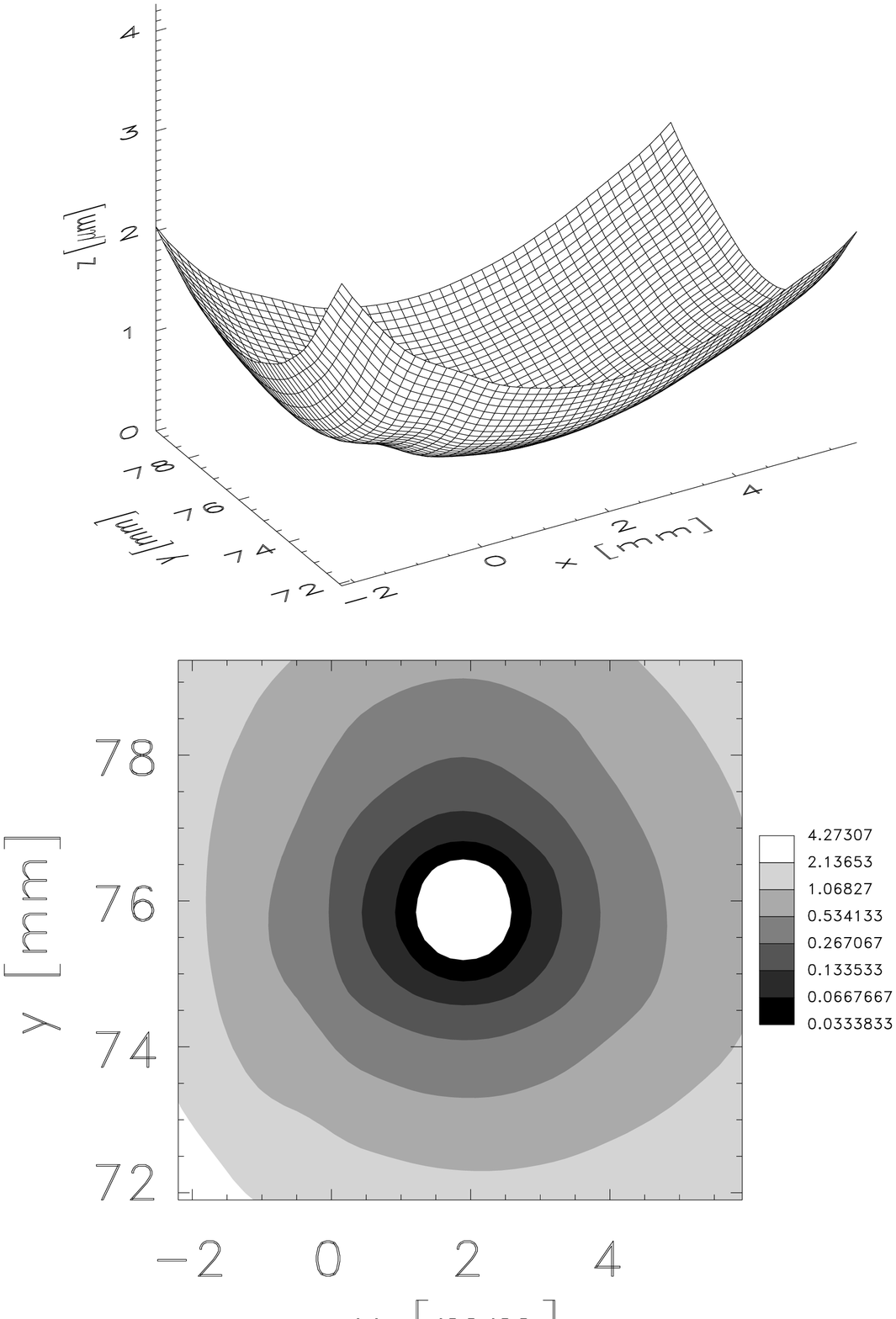}
    \caption{Illustration of the bent plate in real-space. The bending radius is about $14$\,m. The upper part corresponds to region~C, the lower part represents region~B. \label{fig:scheibe}}
\end{figure}

The intrinsic mosaicity of the platelet could not be determined within
these measurements because the platelet turned out to be bent.  Due to
the relatively large width of the beam spot, the signal results from
an overlap of different regions in the bent sample, which enhances the
width of the transverse scans.  Consequently, the width of the
transverse scans at a given spot is a measure of the
bending radius of the plate and not a measure of the intrinsic
mosaicity.  A more accurate determination of the bending radius was
achieved by using a narrow cross-section of the beam, e.g.~$50\times
50$\mue$^2$, and measuring the shift of the position of a main
reflection depending on the position in real space on the plate, which
was oriented perpendicular to the beam.  Using this technique, a
real-space picture of the plate can be reconstructed from the data
[Fig.~\ref{fig:scheibe}].  The bending of the lattice planes is almost
spherical, the bending radius results to $\sim 14$\,m.  In this
figure, region~C corresponds to the upper, concave side, i.e.~the
lower, convex side corresponds to region~B.  Using an optical
microscope it could be seen that not only the lattice planes are bent
but the platelet itself is bent macroscopically.  It should be mentioned that
the bending has been observed not directly after the polishing process
but after the first low temperature measurements, i.e. after the
platelet had undergone the structural phase transition.  Thus it is
not clear if the bending process took place directly after the
polishing process or some weeks later after cooling through the
transition.

\begin{figure}[h]
  \centering \includegraphics[angle=-90,width=\linewidth]{\bilder
    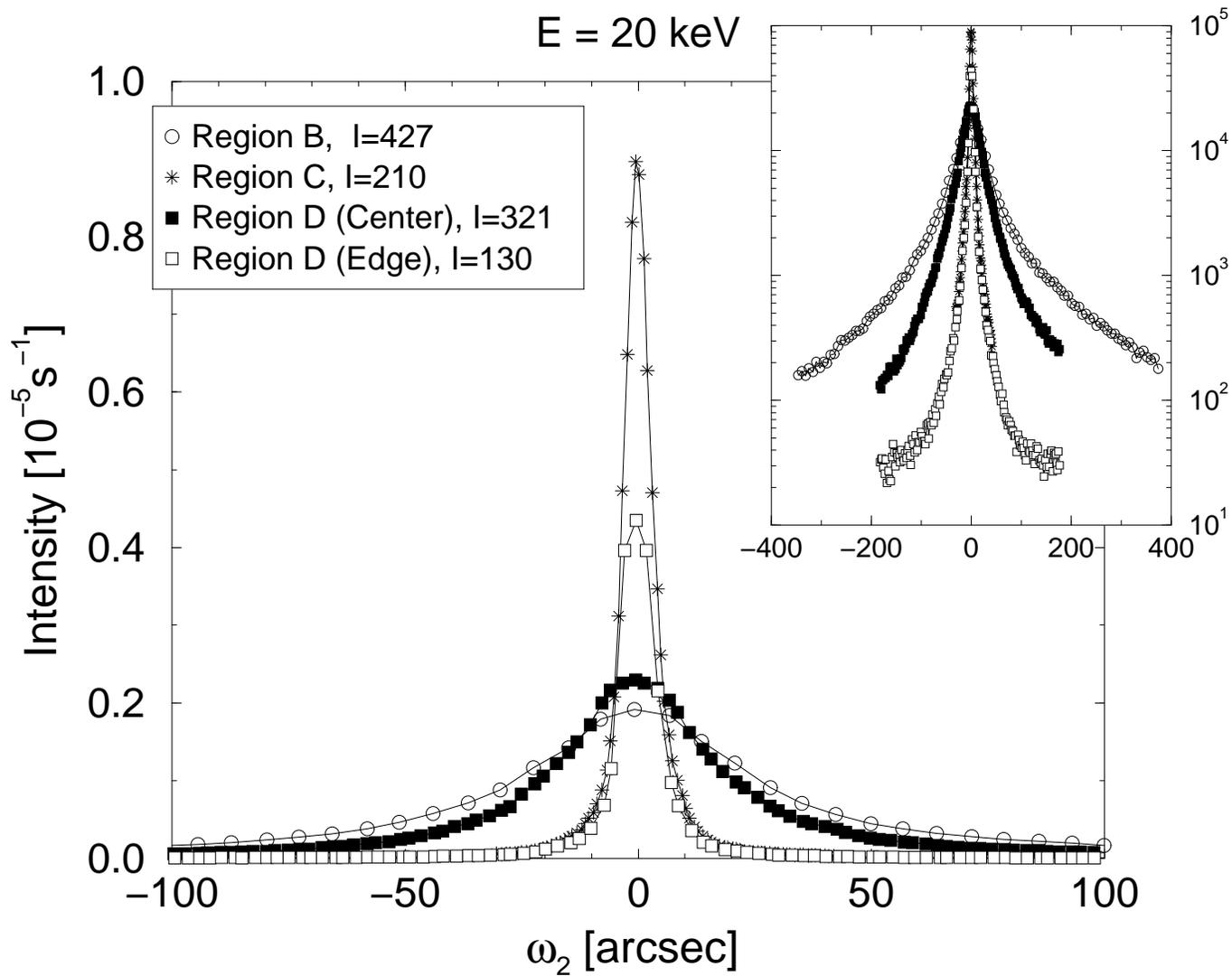}
    \caption{Rocking curves of the ($200$) bragg reflection at the different surfaces of the \shirane-samples, measured at room temperature with $20$\,keV x-rays at beamline D4. The Rocking curves in region~B and D are much broader than those of region~C and of the edge region of the residual block. The inset shows the same data in a logarithmic scale. $I=\dots$ represent the values of the integrated intensities of the respective scans. \label{fig:d4}}
\end{figure}

Supplementary to the depth dependent measurements with $100$\,keV
photons, the surfaces B--D have been investigated in Bragg geometry
with a photon energy of $20$\,keV on a triple-axis diffractometer at
the HASYLAB beamline D4 \cite{Als86}.  The absorption length at this
energy was determined to $\mu^{-1}\sim 55$\,\mue, i.e.~the relevant
contribution to the Bragg peaks results from the surface near region
of some ten microns thickness.  In figure~\ref{fig:d4} the scattering
profiles of the ($200$) Bragg reflection are shown for the three
surfaces B,C and D both in linear scale and in logarithmic scale
(inset).  At the surface of the residual block (region~D) a difference
between the center of the surface and the edge region of the surface
was found.  It can be seen, that regions B and D (center) show much
broader Bragg peaks than region~C and the edge of the residual block
(D edge).  Additionally, the integrated intensity of the respective
scans is shown in the legend.  Consistent with the observations
described above, the values are increased at surfaces B ($I=427$) and
D ($I=321$), compared to surface C ($I=210$).  However, interestingly
the edge region of the residual block shows a similar behaviour as the
cut surface of the plate (region~C).

The characterisation of the crystallographic quantities in sample~I
can be summarised as follows: The residual \shirane{} block consists
of an almost perfect bulk (region~E) surrounded by a layer (region~D)
of increased mosaicity and increased lattice parameter variations.
Both mosaicity and lattice parameter variation increase exponentially
with an 1/e-length of $26(1)$\,\mue{} approaching the surface.  This
behaviour is identical to the depth dependence of the experimental
data from the original sample before the cut (region~A), presented in
\cite{Rue97a}, and supports the interpretations of the earlier
$\gamma$-ray diffraction experiments on the same sample in
\cite{Sch86}.  However, the region~D close to the surface of the
residual block, which has been inside the almost perfect bulk of the
sample before the cut, strongly changed its crystallographic
properties after cutting and polishing the surface.  On the other
hand, the cut surface of the residual plate, i.e.~region~C shows no
effects in the mosaicity or the integrated reflectivity.  The values
are the same as those of the bulk (region~E), only a slightly enhanced
value of the lattice parameter variations is observed.  In spite of
the identical treatment of surfaces~C and D, large differences are
observed. The behaviour of the untreated surface of the
plate, region~B, remains unchanged compared to the earlier
measurements at region~A.

The only explanation for the observed difference in the behaviour of
regions~C and D can be given by the fact that the plate became bent
after the cut.  Assuming that the original sample included long range
strain fields inside the bulk, possibly introduced by the growth
process, a relaxation of these strain fields is not possible in the
residual block because of its dimensions and the high crystallographic
perfection.  In region~D, close to the surface of the residual block,
these strain fields lead to an increased mosaicity and increased
lattice parameter variations in the center of the surface, but not in
the edge regions of the surface of the residual block because here the
strain fields are able to relax. Similarly the relaxation of the
strain fields is observed in the platelet.  The existing strain
gradient in the platelet, experimentally determined at region~A in
\cite{Rue97a}, is able to relax by bending the whole thin sample.
Hence, in region~C no dislocations like in region~D are observed, the
crystallographic parameters remain unchanged compared to the bulk
behaviour.  However, the other surface, region~B, still is full of
dislocations and thus also does not change its general behaviour.  Due
to the large dimensions of the crystal, this bending process of course
is not possible at the surface of the residual block, in region~D.

Similar measurements have been carried out on sample~II, it can be
seen that the surface of sample~II qualitatively shows the same
behaviour like the regions~A, B and D of sample~I.  The mosaicity of
sample~II is about one order of magnitude larger than that of
sample~I, but it is still much better than the mosaic spreads of
30-100'' of the Verneuil-grown crystals.  The depth dependence of the
crystallographic parameters of the Verneuil-grown crystals has not
been investigated.
\section{Critical temperatures}
\label{sec:tc}
\begin{figure}[h]
  \centering \includegraphics[angle=-90,width=\linewidth]{\bilder
    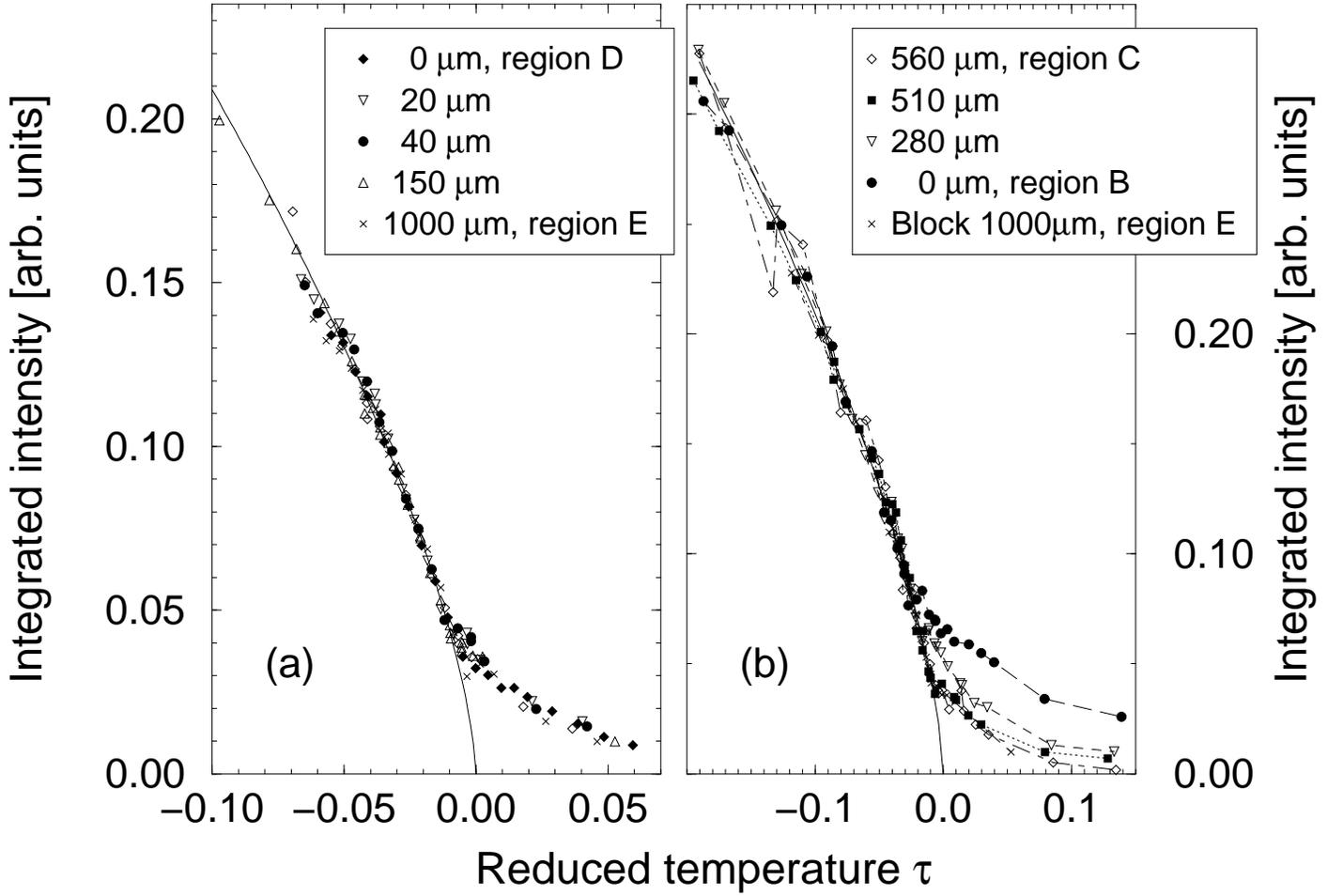}
    \caption{Temperature dependence of the integrated intensities of the ($511$)$/2$ superlattice reflection, normalized to the respective extrapolated value at zero temperature, for different positions (a) in the residual block and (b) the plate. The tail above \tc{} is almost independent of the distance from the surface. Only in region~B significant changes in the temperature dependence of the tail are observed. \label{fig:tails}}
\end{figure}

Essential for the discussion of critical behaviour and the
determination of critical exponents is the accurate measurement of the
critical temperature.  Following the procedure of Riste et
al.~\cite{Ris71}, the temperature dependence of the integrated
intensity of a superlattice reflection has been measured both below
and above the phase transition.  The scattered intensity at this
position is proportional to the square of the order parameter,
$I\propto \langle\varphi\rangle^2 =
\langle\varphi_0+\delta\varphi\rangle^2$.  Unfortunately, just below
\tc{} both the critical fluctuations $\langle \varphi^2\rangle$ and
the static part $\langle \varphi_0\rangle^2$ contribute to the signal.
Above \tc{} the critical fluctuations give rise to a tail in the
temperature dependence of the order parameter
[figure~\ref{fig:tails}].  As described in paper~I, we have neglected
the contribution of static order parameter clusters, which do also
exist above \tc{}.  Using the Landau approximation, the
susceptibilitiy at $T_c-\Delta T$ is a factor of $2$ smaller than the
susceptibility at $T_c+\Delta T$ [see e.g.~\cite{Cow80a}].  This
relation is used to substract the contribution of the fluctuations
below \tc{} as described in \cite{Ris71}.  The temperature dependence
of the residual data $I'$ can be fitted with a power law:
\begin{equation}
  \label{eq:powerlaw}
  I' \propto \langle \varphi_0\rangle^2 \propto \left(\frac{T-T_c}{T_c}\right)^{2\beta}
\end{equation}
In order to maximise the compatibility of the critical temperatures
\tc, the critical exponent $\beta$ was fixed to $\beta=0.34$
\cite{Ris71}.  Slight changes in $\beta$ were allowed to determine the
error bars of the critical temperatures.  The results for the critical
temperature in the bulk of the various samples are summarised in
table~\ref{tab:samples}.  The value $T_c=98.8(2)$\,K for sample~I has
been determined in the bulk of the original sample.  Usually, in 
literature, the critical temperature of the structural phase
transition in \srt{} averages to about $105$\,K.  This value is
similar to the value for the two Verneuil samples with low oxygen
vacancy concentrations.  It has been shown earlier, that an increased
amount of oxygen vacancies reduces the critical temperature
\cite{Bae78,Has78}, the reduced Verneuil sample nicely reproduces the
results.  However, the shift in the critical temperature for the two
highly perfect samples~I and II is much too large compared to the
experimentally determined concentration of oxygen vacancies.  The
origin of this observation is not clear yet, perhaps the high degree
of perfection of samples~I and II is responsible for the low
transition temperature.  As pointed out by Zhong et al.~\cite{Zho96},
the existence of quantum fluctuations leads to a significant decrease
of the critical temperature of about $20$\,K.  It might be speculated
that the quantum fluctuations show a stronger effect in the more
perfect samples than in the worse Verneuil samples, which leads to an
additional decrease of the phase transition temperature.

\begin{figure}[h]
  \centering \includegraphics[angle=-90,width=\linewidth]{\bilder
    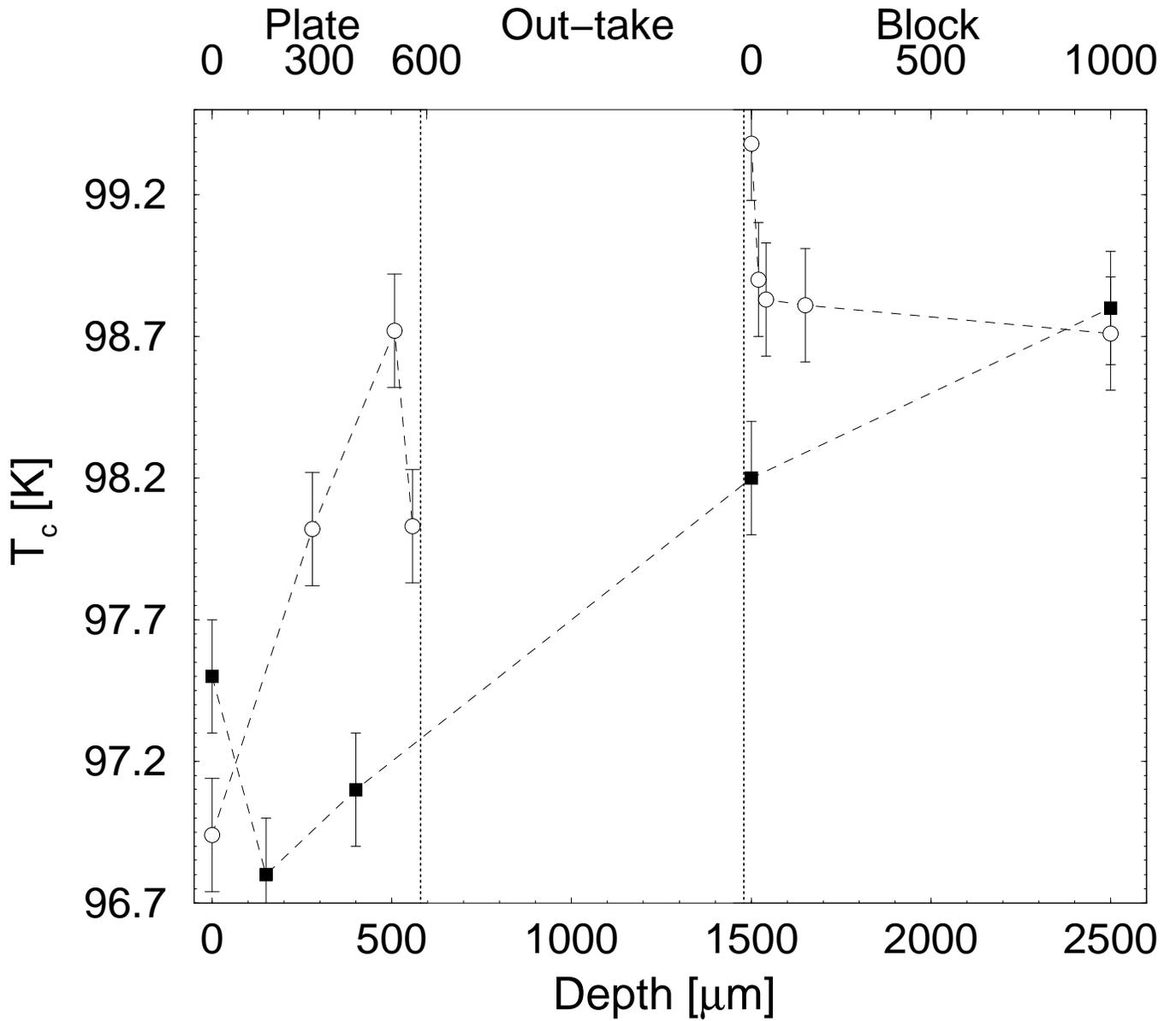}
    \caption{Depth dependence of the critical temperatures in sample~I after the cut (opaque circles), compared to the values for the original sample (black squares). \label{fig:tc}}
\end{figure}

In analogy to the crystallographic characterisations described above,
the depth dependence of the critical temperature has been investigated
in detail in sample~I.  The results are plotted in
figure~\ref{fig:tc}.  Both the transition temperatures before and
after the cut are shown.  As to be expected no changes in \tc{} are
observed in region~E, the bulk of the sample.  In the original sample
two features can be identified: Over a large range the critical
temperature is decreasing and close to the surface, in region~A, a
slightly enhanced value for \tc{} is found.  The latter observation is
similar to the behaviour at the surface of the residual block,
region~D.  There also a slight increase in the critical temperature is
found very close to the surface ($\sim 20$\,\mue).  However, in the
residual plate the general trend is unclear.  $50$\,\mue below the cut 
surface (region~C) of the plate, \tc{} is identical to
the bulk value (region~E), but at both surfaces, regions~B and C, the
critical temperature decreases substantially.  Comparing these
observations with the results of the crystallographic
characterisation, it can be concluded that two different effects have
to be distinguished.  On one hand, the probably large amount of
dislocations and other defects, which are not yet identified, in
region~A is responsible for the large difference in the transition
temperatures found in the bulk and near the surface of the original
block.  On the other hand, it seems that the strain fields in its
surface near region increase the transition temperatures slightly.
This is observed in region~A and region~D, but not in region~C, where
the strain fields are absent.  After the relaxation of the strain
gradient in the plate, the critical temperature in region~C is close
to the bulk value (region~E).  In region~B \tc{} strongly declines
almost to the low value observed at the original surface (region~A).

Another interesting result of the measurements is the behaviour of the
tails of the order parameter above the critical temperatures.  In
order to compare the tails for the different samples, the integrated
intensities of the superlattice reflection have been normalised to the
respective scaling factor of the fitted power law, i.e.~the
extrapolated value at zero temperature.  Also, the temperature is
replaced by the reduced temperature $\tau=\frac{T-T_c}{T_c}$.  Using
this method, the effect of the stronger, more kinematical scattering
at the surface [see section~\ref{sec:sam}] is cancelled out and all
curves except one coincide below the transition temperature, as can be
seen in figure~\ref{fig:tails}.  The additional information derived
from this kind of presentation is the similarity of the temperature
dependence of the tails above the critical temperatures.  In the
residual block [fig.~\ref{fig:tails}a], all data points follow the
same trend above \tc, i.e.~ the nature of the tails is not connected
to the observed strain gradients at the surface.  Moreover, the tails
are identical in the residual block and in the residual plate
[fig.~\ref{fig:tails}b], only the old surface of the plate, region~B,
shows a significantly enhanced amount of scattering above \tc{}.
Certainly, this has to be attributed to the large amount of
dislocations and/or other defects in this region.
\section{The sharp component in the critical scattering}
\label{sec:crit}
\begin{figure}[h]
  \centering \includegraphics[angle=-90,width=\linewidth]{\bilder
    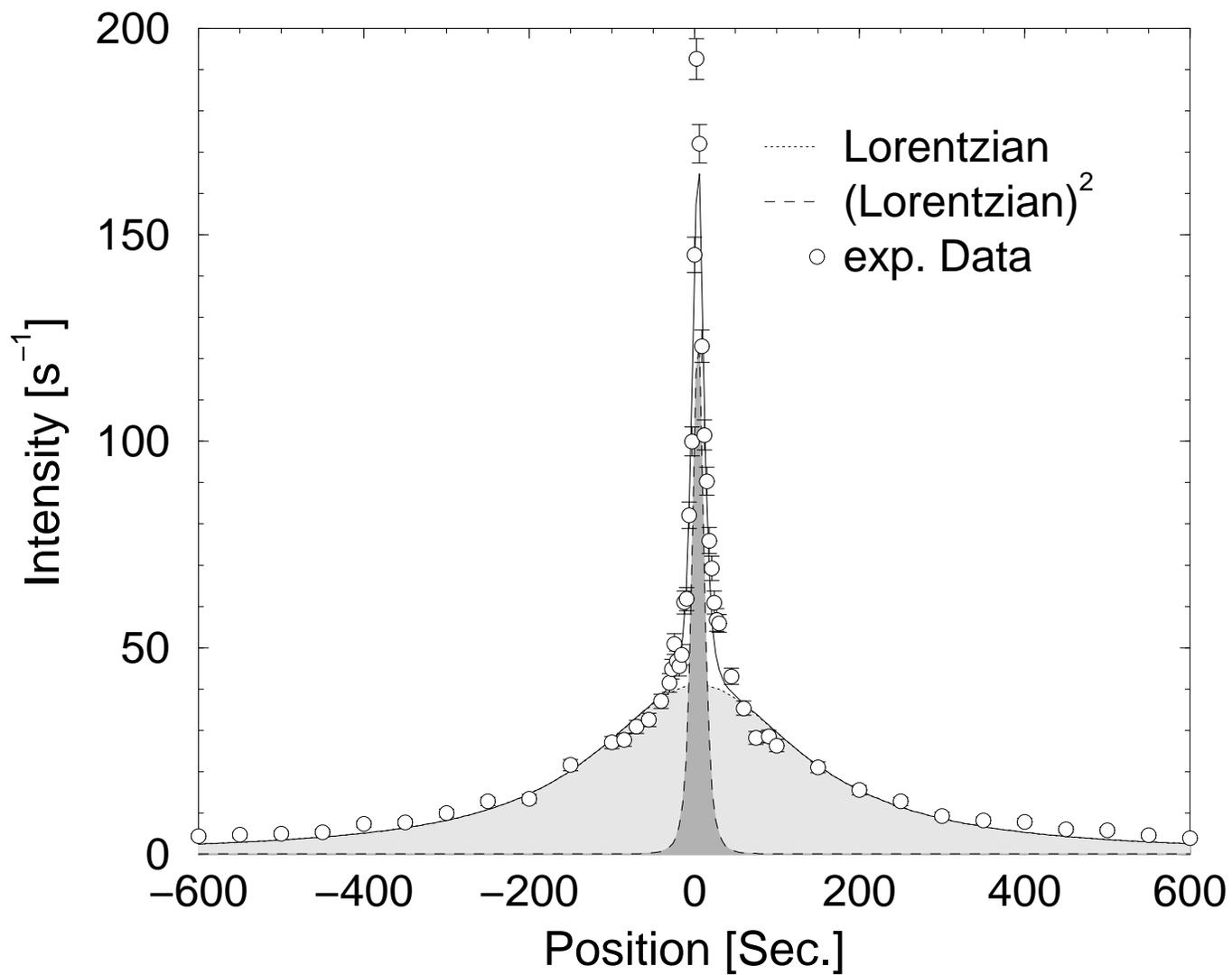}
    \caption{Transverse scan $0.5$\,K above \tc{} in a distance of $20$\,\mue{} from the surface of the residual block (region~D). Additional to the broad Lorentzian distribution a sharp Lorentzian squared profile is visible at the position of the superlattice reflection. \label{fig:typical}}
\end{figure}

The critical scattering has been observed at the position of the
($511$)/2 superlattice reflection in the samples~I and II and at the
position of the ($531$)/2 superlattice reflection in the Verneuil
samples.  In most cases, after deconvolution of the experimental
resolution [see appendix~\ref{sec:app}] the scattering profile could
be fitted with a single Lorentzian distribution function and as a
result the inverse correlation length $\kappa_{Lor}$ and the
susceptibility $\chi_{Lor}$ of the broad component in the critical
scattering could be extracted.  Plotting this as a function of
temperature it is possible to deduce the values of the critical
exponents \nb{} and \gb{}, which has been discussed in detail in
paper~I.  In this paper, we want to focus on the sharp component,
which has been observed only in samples~I and II in surface near
regions.  In those cases the scattering profile had to be fitted with
a sum of a Lorentzian and a Lorentzian-squared profile, i.e.~two
additional parameters, the inverse correlation length $\sigma_{Lq}$
and the susceptibility $\chi_{Lq}$, have to be taken into account (the
peak-position of the two contributions turned out to be identical).  A
typical example of a measured scattering profile is shown in
figure~\ref{fig:typical}.  The Lorentzian-squared part is much
narrower than the Lorentzian contribution, which is the reason for
calling it the sharp component.

\begin{figure}[h]
  \centering \includegraphics[angle=-90,width=\linewidth]{\bilder
    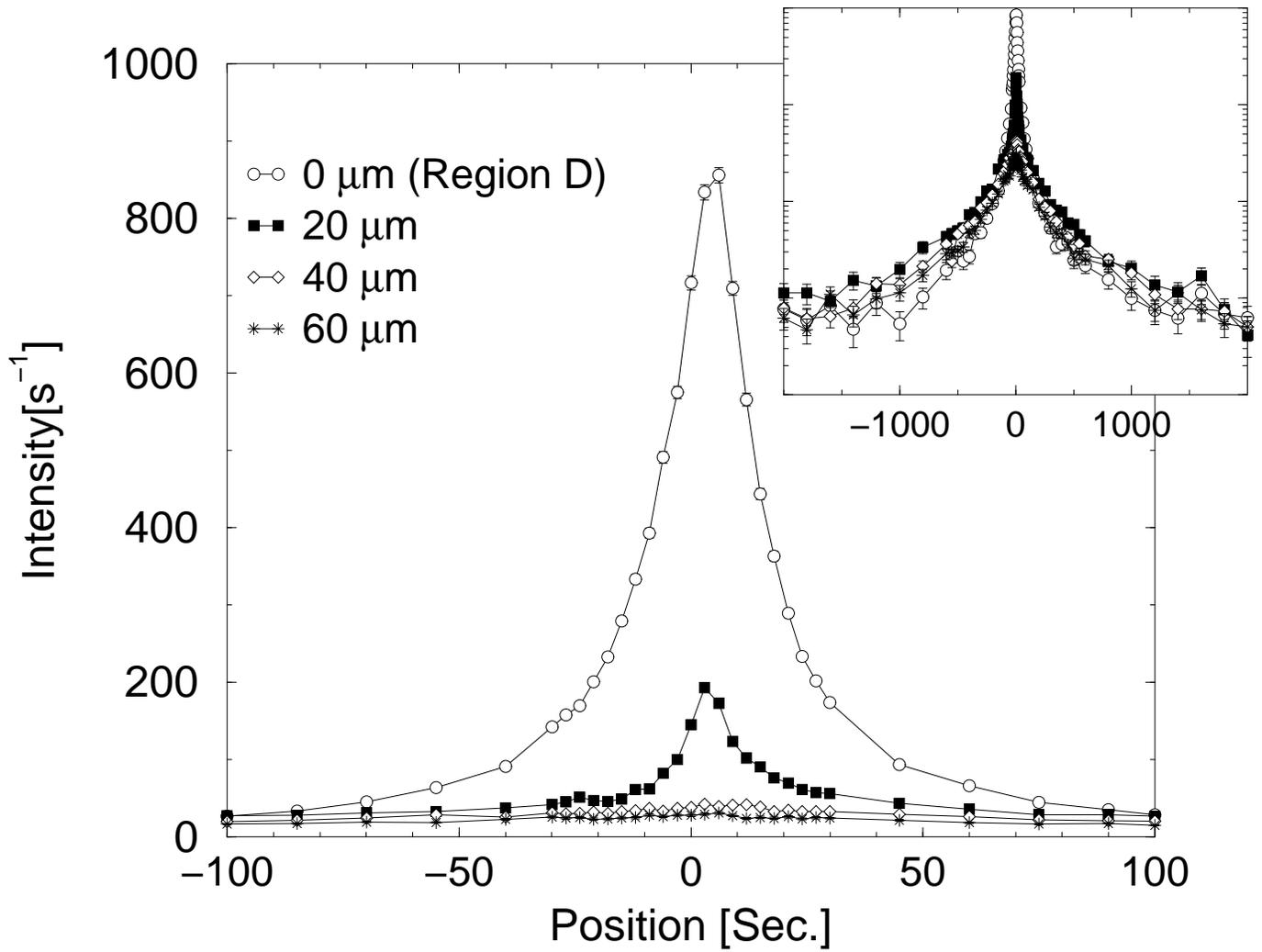}
    \caption{Transverse scan profile of the ($511$)$/2$ superlattice reflection about $1$\,K above the critical temperature for different distances to the surface of the residual \shirane-block (sample~I, region~D). For the top $20$\,\mue{} the intensity is strongly enhanced in a narrow region around the superlattice position.
      The inset shows the measured data over a much larger angular
      range in a logarithmic scale. It can be seen that without
      considering the sharp component all profiles are almost
      identical. \label{fig:depth_block}}
\end{figure}

At a temperature about $1$\,K above \tc{} the depth dependence of the
critical scattering profile in the residual block of sample~I is shown
in figure~\ref{fig:depth_block}.  In the top $20$\,\mue{},
corresponding to region~D, the additional sharp component is clearly
visible.  In the inset, the same data are plotted over a wider angular
range in logarithmic scale.  This is to stress the observation that
the broad component is independent of the location of the probed
volume element in the sample.  The scattering only differs in the
central region, where the sharp component dominates.  The temperature
dependence of the inverse correlation length for the broad component
was identical at all investigated positions in sample~I after the cut.
The values \nb{} and \gb{} for the critical exponents of the broad
component in the different samples and for the various positions in
the samples are summarised in table~\ref{tab:broad}.  As discussed in
paper~I, the absolute values depend on the concentration of the oxygen
vacancies.  However, all values of the critical exponents related to
the broad component can be explained in terms of the influence of
order parameter clusters, the universal behaviour is not restricted.
For the different positions in sample~I the critical exponents for the
broad component are almost identical, i.e.~the occurrence of the sharp
component as well as the increased mosaicity and lattice parameter
variations do not affect the behaviour of the broad component.  In
table~\ref{tab:broad} also the validity of some scaling relations is
checked ($\alpha$, $\beta$, $\gamma$, $\nu$ and $\eta$ are the
well-known critical exponents, $d$ is the dimensionality of the
system, in this case $d=3$).
\begin{eqnarray}
  \gamma &=&(2-\eta)\nu\label{eq:scale1}\\
  2-\alpha&=&\gamma+2\beta\nonumber\\
  2-\alpha&=&d\nu\nonumber\\
\Rightarrow \beta&=&\frac{1}{2}(d\nu-\gamma)\label{eq:scale2}
\end{eqnarray}
The value of $\eta$ is very small ($\sim 0.03$) for \srt{}
\cite{Fis64}, i.e.~the ratio \gdnb{} should be close to $2$.
Especially the Verneuil-samples almost perfectly fulfill this
condition, overall this scaling relation holds quite nicely in all
cases.  Using relation~(\ref{eq:scale2}), the obtained value of
$\beta$ might be compared with the results below \tc{}, where $\beta$
was fixed to the value $0.34$.  In the surface near regions~B and D
this scaling relation seems to fail, but with respect to the large
error bars the agreement is satisfying for most of the samples.

\begin{figure}[h]
  \centering \includegraphics[angle=-90,width=\linewidth]{\bilder
    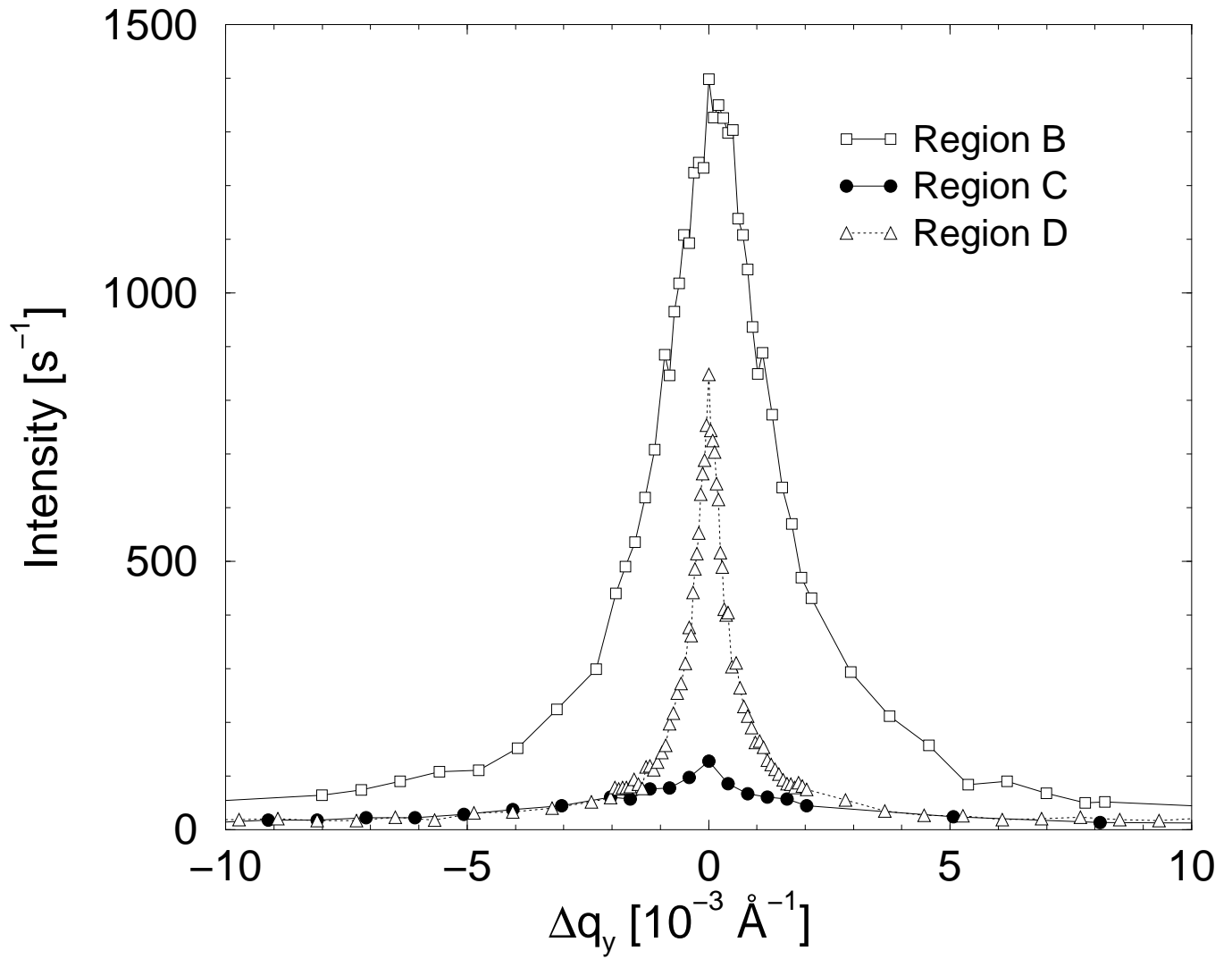}
    \caption{Comparison of the scattering profiles of the ($511$)$/2$ superlattice reflections of sample~I $1$\,K above \tc{} measured close to the three different surfaces. The old surface of the plate (region~B) differs strongly from the other ones, whereas the new surface of the plate (region~C) and the residual block (region~D) differ only in the narrow region in the center of the plot. The broad component is identical for the latter two surfaces, but at the residual block the sharp component can be observed additionally. \label{fig:shirane_comp}}
\end{figure}

\begin{figure}[h]
  \centering \includegraphics[angle=-90,width=\linewidth]{\bilder
    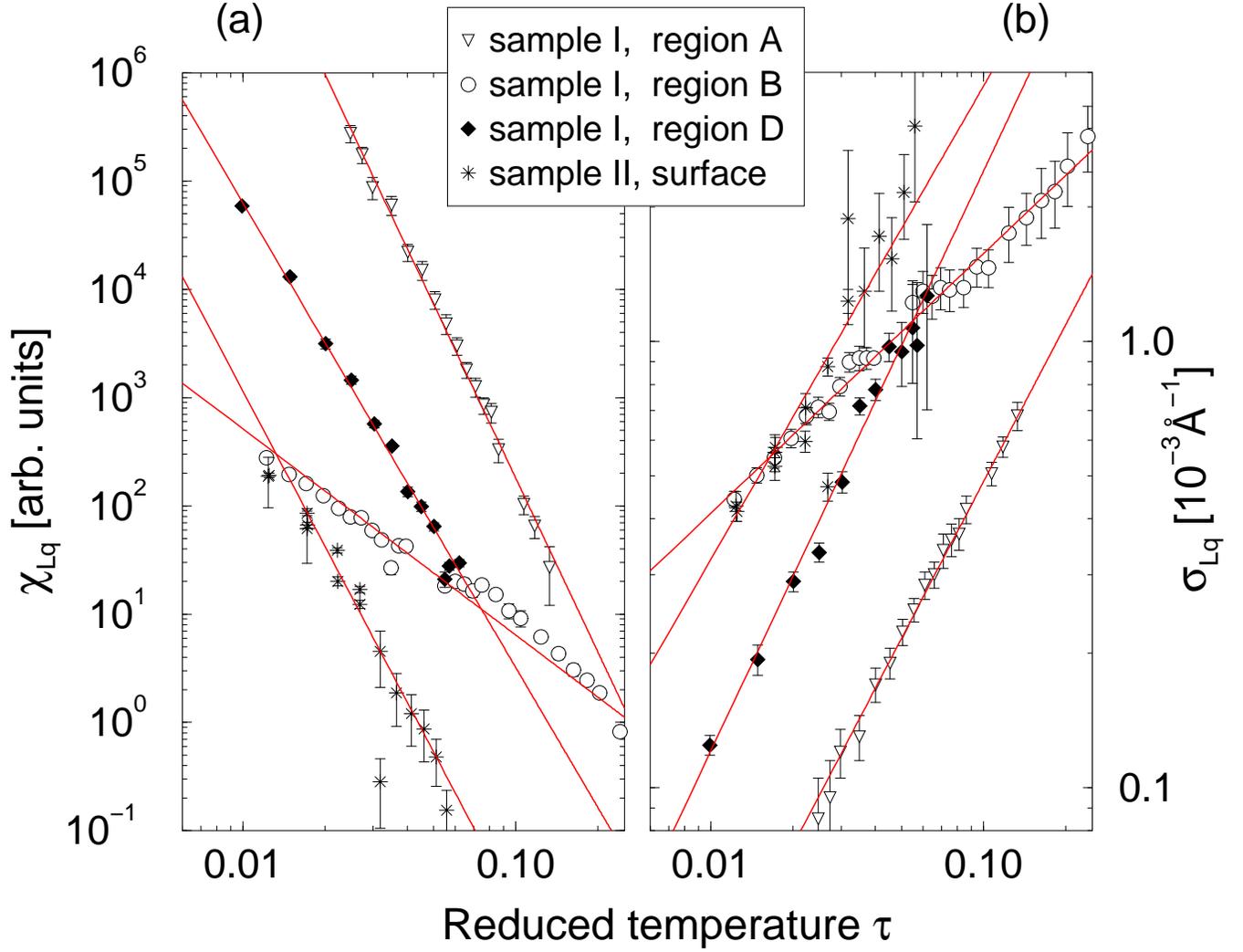}
    \caption{Temperature dependence (a) of the inverse correlation length $\sigma_{Lq}$ and (b) of the  susceptibility $\chi_{Lq}$ of the sharp component for the different surfaces. The critical exponents $\nu_s$ and $\gamma_s$ are similar for most of the investigated surfaces, but they strongly deviate at the old surface of the plate (region~C). \label{fig:crit_sharp}}
\end{figure}

The dircet comparison between the three surfaces of the residual
\shirane-sample, i.e.~regions B, C and D is shown in
figure~\ref{fig:shirane_comp}.  Apparently, the width of the sharp
component in region~B is much broader than the respective width in
region~D.  Furthermore, almost no signal of the sharp component is
visible in region~C.  The profiles for regions~C and D are identical
except for a narrow region in the center of the scan, which is due to
the sharp component.  In figure~\ref{fig:crit_sharp} the plots from
which the critical exponents \gs{} [fig.~\ref{fig:crit_sharp}a] and
\ns{} [fig.~\ref{fig:crit_sharp}b] have been determined are shown for
the four positions where the sharp component was observed.  Again, the
old surface of the cut-off plate, region~B, strongly differs from all
other data sets.  The resulting values for the critical exponents of
the sharp component are listed in table~\ref{tab:sharp}.  The strong
deviation of the behaviour in region~B is obvious, the second length
scale in the three other regions is described by similar critical
exponents.  Interestingly, the ratio \gdns{} is noticeable larger than
$2$, more likely close to $4$.  The validity of scaling relations for
the sharp component could be established, if the intensity of the
sharp component is not proportional to the susceptibility but to the
square of the susceptibility, as it is also the case e.g.~for
Huang-scattering \cite{Schw91}.  The two functions
\begin{equation}
\label{eq:lorentzsquared_alt}
  I_{Lq}  =\frac{\chi_{Lq}}{\left(1+\left(\frac{\qvec}{\sigma_{Lq}}\right)^2\right)^2} = \hat{\chi}_{Lq}(\qvec,T)
\end{equation}
and
\begin{equation}
\label{eq:lorentzsquared_neu}
  I_{Lq}  =\left(\frac{\chi_{Lq}'}{1+\left(\frac{\qvec}{\sigma_{Lq}}\right)^2}\right)^2 = \hat{\chi'}_{Lor}^2(\qvec,T)
\end{equation}
only differ by the definition of the respective susceptibility.  All
critical exponents discussed so far have been derived using
equation~(\ref{eq:lorentzsquared_alt}).  However, assuming a
Lorentzian distribution and using the latter definition, the
susceptibility $\chi_{Lq}$ is replaced by
$\chi_{Lq}'=\sqrt{\chi_{Lq}}$.  As a result the value of the critical
exponent is reduced by a factor of two, $\gamma_s'=\gamma_s/2$ and the
ratio $\gamma_s'/\nu_s$ is close to two, fulfilling the scaling
relation~(\ref{eq:scale1}).

Concluding, the sharp component in the critical scattering in \srt{}
occurs in surface near regions (some ten microns) of highly perfect
crystals, but not if long range strain fields are absent in this
region.  The temperature dependence of the correlation length and the
susceptibility can be described with critical exponents, which do
fulfill scaling relations if a definition of the scattering profile
different from current convention is applied.  The occurrence of the
sharp component does not affect the critical behaviour of the broad
component in any manner.
\section{Discussion}
\label{sec:sum}
We have investigated the critical scattering above the $105$\,K
structural phase transition of \srt.  The main focus is put on the
dependence of the critical phenomena on the depth below the surface.
In order to interpret the results concerning the critical scattering,
the crystallographic perfection was analysed with high momentum- and
real-space resolution especially in the regions close to the surfaces.
Particularly we concentrated on the differences of several surfaces of
a highly perfect sample.

As a result it is observed that one necessary condition for the
existence of two length scales in the critical scattering is the
vicinity (some ten microns) of the surface of a sample.  Another
necessary condition is the existence of strain fields in this region,
experimentally demonstrated by the increase of lattice parameter
variations and mosaicity close to the surface.  However, in samples of
lower quality, i.e.~with strongly enhanced mosaic spreads the sharp
component is suppressed significantly compared to that found in the
highly perfect crystals.  Possibly the long range order of the
critical fluctuations is destroyed by the high amount of topological
defects in samples of very high mosaicity.

Starting with the bulk of sample~I, region~E, no sharp component can
be observed, because it is not a surface near region.  In region~D,
both conditions are fulfilled.  Both mosaicity and the variation of
the lattice parameter, as well as the critical temperature, increase
exponentially approaching the surface and the exisitence of long range
strain fields at this surface seems reasonable.  Simultaneously, also
the intensity of the sharp component increases, while the broad
component remains unaffected.  Region~C consists of an almost perfect
crystal, where the strain fields were able to relax by bending of the
lattice planes.  Thus, no sharp component is observed and the critical
temperature is close to the bulk value.  On the other hand, region~B,
the ``old'' surface of the plate, still is full of dislocations and
probably other unspecified defects.  Here, both conditions for the
appearence of the sharp component are fulfilled, but different from
the surface of the residual block, region~D, a lot of dislocations
exist at this surface stemming from the frequent use over more than
twenty years in many experiments.  Moreover, the bending of the
lattice planes additionally deteriorated the crystallographic
perfection, because unlike the other almost perfect surface no
relaxation was possible at this surface.  The development of long
range correlations at this surface is suppressed or, at least,
affected.  The result is the large deviation in the values of the
critical exponents for the sharp component in region~B, compared to
the earlier results on the same surface before the cut, called
region~A.  The behaviour of sample~II at the surface is very similar
to the behaviour of sample~I in region~D.  The surface of sample~II
was cut, polished and etched, as described in paper~I.  Afterwards in
the surface near region mosaicity, lattice parameter variation and the
intensity of the sharp component increased simultaneously.  Compared
to the two almost perfect samples the intensity of the sharp component
in the Verneuil-grown samples III-V is suppressed by about three
orders of magnitude, i.e.~no evaluation of critical exponents for this
component was possible.  This is due to the large mosaicity, which
suppresses the long range correlation comparable to the effect at the
old surface of the plate (region~B).  Qualitatively, it has been shown
before that at distorted surfaces the intensity of the sharp component
is reduced \cite{Thu94,Wat96}.

In conclusion, long range strain fields in the vicinity of the surface
seem to be responsible for the second length scale in the critical
scattering of \srt.  These long range strain fields can spread out
much better in almost perfect crystals, dislocations and other
topological defects reduce the strain fields and affect the critical
exponents and eventually lead to complete suppression of the second
length scale.  Similar to the usual critical fluctuations, the
temperature dependence of the long range correlations can be described
with a set of critical exponents, also we presented indications for
the validity of scaling laws.  However, the distribution of defects in
the surface near region is crucial for the formation of the sharp
component and for the values of the critical exponents describing this
phenomenon.  This interpretation of the nature of the sharp component
is consistent with the ideas of Cowley \cite{Cow96a}, who suggested
that at the surface the coupling of strain fields to the order
parameter might lead to free surface waves with a higher effective
transition temperature than the bulk fluctuations.  The sharp
component is then attributed to the free surface fluctuations.
Unfortunately, no quantitative calculations on the basis of this
approach have been performed yet.  Further theoretical work is needed
for a full quantitative understanding of the nature of the second
length scale in the critical scattering.
\section*{Acknowledgements}
We would like to thank H.J.~Scheel for making available the flux-grown
SrTiO$_3$ crystal, S.~Kapphan for preparing the Verneuil crystals and
E.~Courtens for stimulating remarks.  Valuable comments from
B.~Kaufmann and F.~Schwabl are gratefully approved.  Support by US
Department of Energy under contract No.~DE-AC02-98CH10886 is
acknowledged.
\appendix
\section{Deconvolution of the resolution function}
\label{sec:app}
\begin{figure}[h]
  \centering \includegraphics[angle=-90,width=\linewidth]{\bilder
    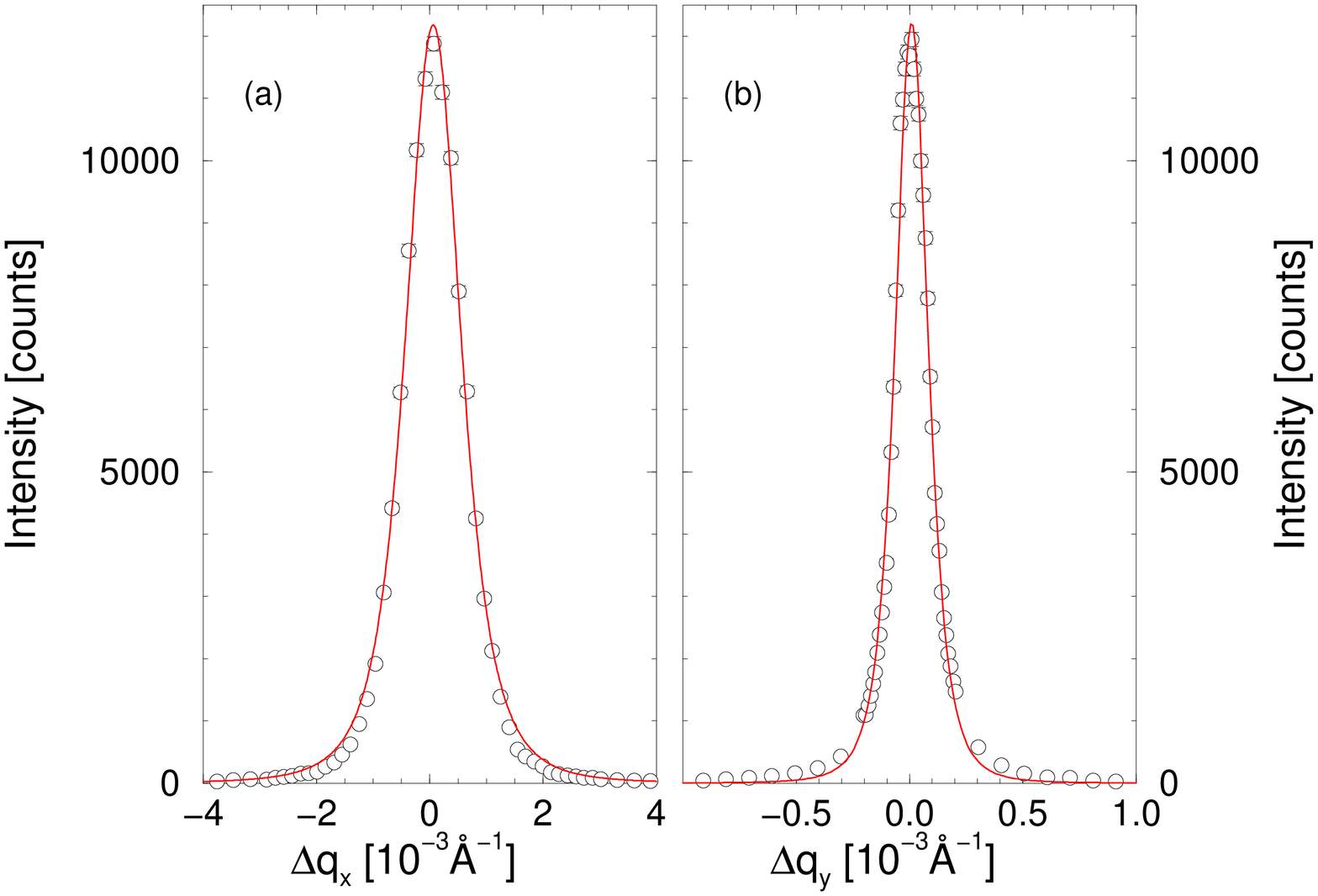}
    \caption{Longitudinal (a) and transverse (b) scattering profiles of the ($511$)/2-superlattice reflection a few degrees below the critical temperature. The solid lines represent the best fits to the data using a Lorentzian-squared profile. \label{fig:resolution}}
\end{figure}

\begin{figure}[h]
  \centering \includegraphics[angle=-90,width=\linewidth]{\bilder
    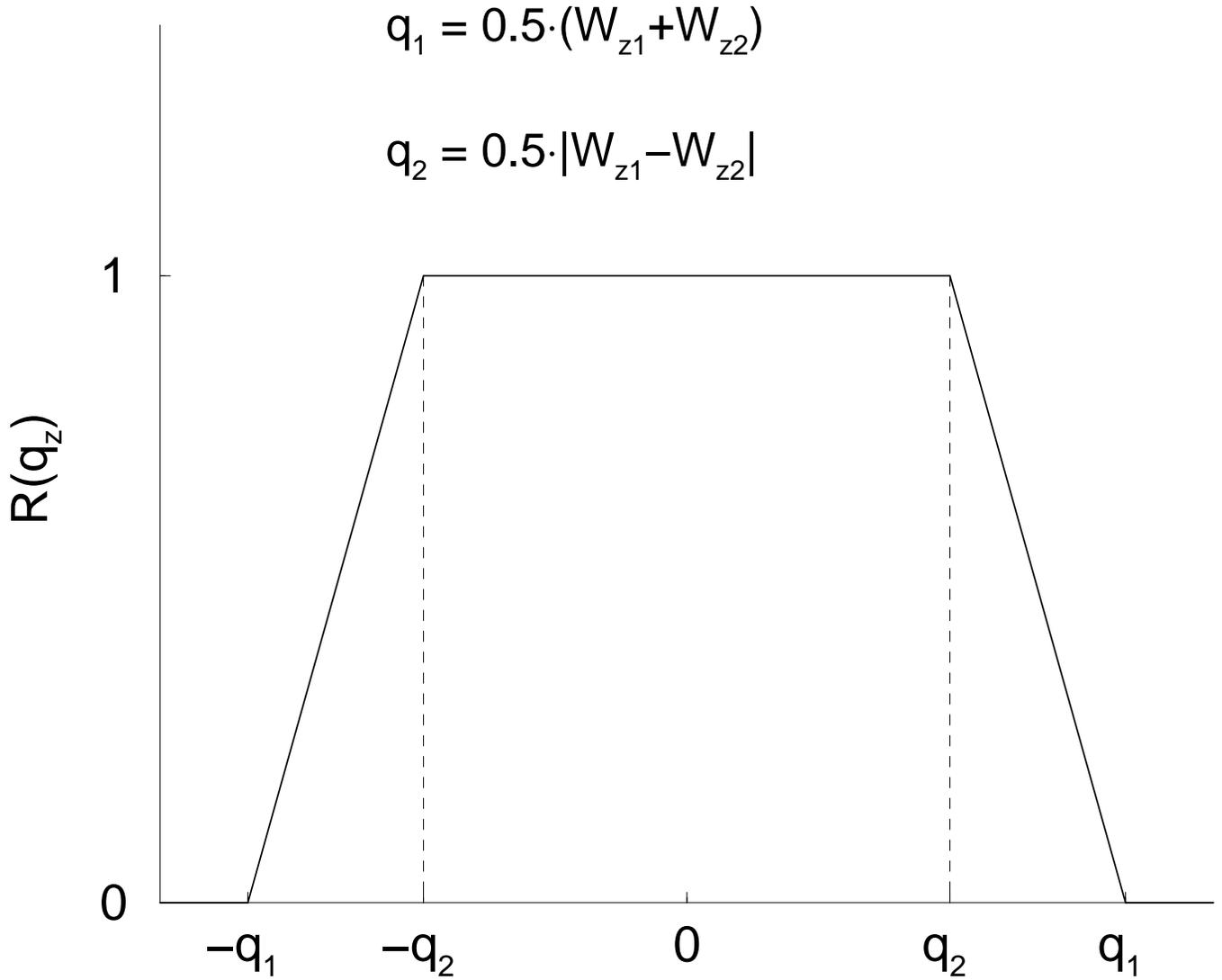}
    \caption{Resolution function perpendicular to the scattering plane. The horizontal part of the trapezoid has a width of $2q_2=|W_{z1}-W_{z2}|$, the width of the sloped parts just corresponds to the width of the smaller slit. In the case $W_{z1}\ll W_{z2}$ the trapezoid degenerates to a rectangle. \label{fig:trapez}}
\end{figure}

In order to extract the correlation length and the susceptibility from
the experimental data, first the experimental resolution function had
to be determined.  The deconvolution procedure is based on the paper
by Hirota et al.~\cite{Hir95a}.  In the scattering plane, the
resolution function was determined experimentally at the position of
the superstructure reflection a few degrees below the transition
temperature [see fig.~\ref{fig:resolution}].  From these scans we
extracted the widths in the longitudinal ($FWHM_x$) and in transvere
($FWHM_y$) directions and the shape of the respective scattering
profiles.  In most cases, the shape of the superstructure peak was a
lorentzian-squared profile.  The functional form of the resolution
element was approximated by:
\begin{eqnarray}
  R(q_x,q_y)&=&\left(1+\left(\frac{q_x}{\sigma_x}\right)^2+\left(\frac{q_y}{\sigma_y}\right)^2\right)^{-2}\label{eq:aufl_exp}\\
\sigma_{x,y}&=&\frac{FWHM_{x,y}}{2\sqrt{\sqrt{2}-1}}\nonumber
\end{eqnarray}
For the investigated samples this approximation is in very good
agreement with the correct resolution function, the theoretical
calculation is explained in detail in \cite{Rue95d,Neu94a}.
Perpendicular to the scattering plane, the resolution is determined by
the opening of the vertical slits, substantially by the widths of the
slits in front of the sample ($S_{z1}$) and in front of the detector
($S_{z2}$).  These widths have to be transformed to the $\qvec$-space
using the distance $L$ between sample and detector and the value of
the wavevector $\kvec$:
\begin{equation}
  \label{eq:qz}
  W_{zi} = |\kvec|\cdot \frac{S_{zi}}{L}
\end{equation}
The resolution element $R(q_z)$ is now given by the convolution of two
rectangular profiles, which is a trapezoidal function, as shown in
figure~\ref{fig:trapez}:
\begin{eqnarray}
  R(q_z)&=&\left\{
      \begin{array}[h]{r@{\quad\forall\quad}r}
0&|q_z|\ge q_1\\
\frac{|q_z|-q_1}{q_2-q_1}&q_2<|q_z|<q_1\\
1&|q_z|\le q_2
      \end{array}\right.  \label{eq:trapez}\\
q_1&=&\frac{1}{2}(W_{z1}+W_{z2})\nonumber\\
q_2&=&\frac{1}{2}|W_{z1}-W_{z2}|\label{trapez2}  
\end{eqnarray}
For an incident intensity $I_0$ the scattered intensity $I(\qvec)$
results to
\begin{equation}
  \label{eq:i_krit_falt}
  \frac{I(\qvec)}{I_0} = \int\int\int dq_x' dq_y' dq_z' R(q_x',q_y')\cdot R(q_z')\cdot I_{crit}(\qvec-\qvecp)\quad,
\end{equation}
where $I_{crit}$ is the intrinsic profile of the critical scattering,
typically a lorentzian distribution.  As described above, in some
cases the critical scattering can not only be described by a single
lorentzian profile.  We used an additional isotropic lorentziansquared
function to fit the experimental data.
\begin{eqnarray}
  I_{crit}&=&I_{Lor}+I_{Lq}\label{eq:i_krit}\\
  I_{Lor} &=&\frac{\chi_{Lor}}{1+\sum_{i=x,y,z}\left(\frac{q_i-q_{i,0}}{\kappa_{Lor,i}}\right)^2}  \label{eq:lorentz}\\
  I_{Lq}  &=&\frac{\chi_{Lq}}{\left(1+\left(\frac{\qvec-\qvecn}{\sigma_{Lq}}\right)^2\right)^2}  \label{eq:lorentzsquared}
\end{eqnarray} 
According to \cite{Shi93}, the lorentzian contribution shows an
anisotropy , the inverse correlation lengths $\kappa_{100}$ in
($100$)-directions are a factor of $1.8$ smaller than the widths
$\kappa_{011}$ in ($011$)-directions.  In our experiments the
scattering geometry was such that $q_x \parallel $ ($511$), $q_y
\parallel $ ($\bar{1}50$) and $q_z \parallel $
($\bar{5}\,\bar{1}\,26$), i.e.~all three directions are slightly
tilted with respect to the main axes.  For simplicity we assumed
$\kappa_x\approx\kappa_y\approx\kappa_z\approx\kappa_{100}\equiv\kappa_{Lor}$,
probably slightly underestimating the correct inverse correlation
length.

The integration in $q_z$ in equation~(\ref{eq:i_krit_falt}) was
performed analytically.  Using the symmetry of the resolution
function, the integral reduces to
\begin{eqnarray}
  \label{eq:i1}
  \int dq_z' R(q_z')\cdot I_{crit}(\qvec-\qvecp) &=&2\int_{0}^{q_2} dq_z' I_{crit}(\qvec-\qvecp)\nonumber\\
&+& 2\int_{q_2}^{q_1} dq_z' \frac{q_z'-q_1}{q_2-q_1} \cdot I_{crit}(\qvec-\qvecp)
\end{eqnarray}
The integral consist of four parts, two addends for both the
lorentzian $I_{Lor}^{conv}$ and the lorentziansquared contribution
$I_{Lq}^{conv}$.  Now we define
\begin{eqnarray}
  b_L^2 &=&\kappa_{Lor}^2+(q_x-q_x')^2+(q_y-q_y')^2\\
  b_{Lq}^2&=&\sigma_{Lq}^2+(q_x-q_x')^2+(q_y-q_y')^2
\end{eqnarray}
and set $q_z=0$, because we are only looking at the intensity in the
scattering plane.  The result for the lorentzian part can be written
as
\begin{eqnarray}
  \label{eq:lor_part}
  I_{Lor}^{conv} &=&\frac{\chi_{Lor}\kappa_{Lor}^2}{b_L}\cdot\left(2\arctan\left(\frac{q_2}{b_L}\right)+\frac{j_1-j_2}{q_2-q_1}\right)
\end{eqnarray}
where
\begin{eqnarray}
j_1&=&b_L\cdot\ln(b_L^2+q_1^2)-2q_1\arctan(q_1/b_L)\\
j_2&=&b_L\cdot\ln(b_L^2+q_2^2)-2q_1\arctan(q_2/b_L)\quad.
\end{eqnarray}
Equation (\ref{eq:lor_part}) can be simplified to:
\begin{eqnarray}
  \label{eq:lor_part_final}
  I_{Lor}^{conv} &=&\frac{\chi_{Lor}\kappa_{Lor}^2}{q_2-q_1}\cdot\left(\ln\left(\frac{b_L^2+q_1^2}{b_L^2+q_2^2}\right)-2\frac{q_1}{b_L}\arctan\left(\frac{q_1}{b_L}\right)+2\frac{q_2}{b_L}\arctan\left(\frac{q_2}{b_L}\right)\right)
\end{eqnarray}
Similarly the integration of the lorentziansquared part yields:
\begin{eqnarray}
  I_{Lq}^{conv} &=&\frac{\chi_{Lq}\sigma_{Lq}^4}{b_{Lq}^3(b_{Lq}^2+q_2^2)}\cdot\left(q_2 b_{Lq}+(q_2^2+b_{Lq}^2)\cdot\arctan\left(\frac{q_2}{b_{Lq}}\right)\right.\nonumber\\
&+&\left. b_{Lq} q_2^2-q_1 b_{Lq}+q_1 (b_{Lq}^2+q_2^2)\cdot\left(\arctan\left(\frac{q_1}{b_{Lq}}\right)-\arctan\left(\frac{q_2}{b_{Lq}}\right)\right)\right)\nonumber\\
\Leftrightarrow I_{Lq}^{conv}&=&\frac{\chi_{Lq}\sigma_{Lq}^4}{b_{Lq}^3(q_2-q_1)}\cdot\left(q_2\arctan\left(\frac{q_2}{b_{Lq}}\right)-q_1\arctan\left(\frac{q_1}{b_{Lq}}\right)\right)
\end{eqnarray}
The remaining calculations,
\begin{eqnarray}
   \frac{I(q_x,q_y)}{I_0} = \int\int dq_x' dq_y'  R(q_x',q_y')\cdot (I_{Lor}^{conv}+I_{Lq}^{conv})\quad,
\end{eqnarray}
have been evaluated numerically, using the approximation
(\ref{eq:aufl_exp}) for the resolution element in the scattering
plane, and were compared to experimental data $I(q_x=0,q_y)$ for
transverse scans or $I(q_x, q_y=0)$ for longitudinal scans in every
step of the fitting procedure.

\begin{figure}[h]
  \centering \includegraphics[angle=-90,width=\linewidth]{\bilder
    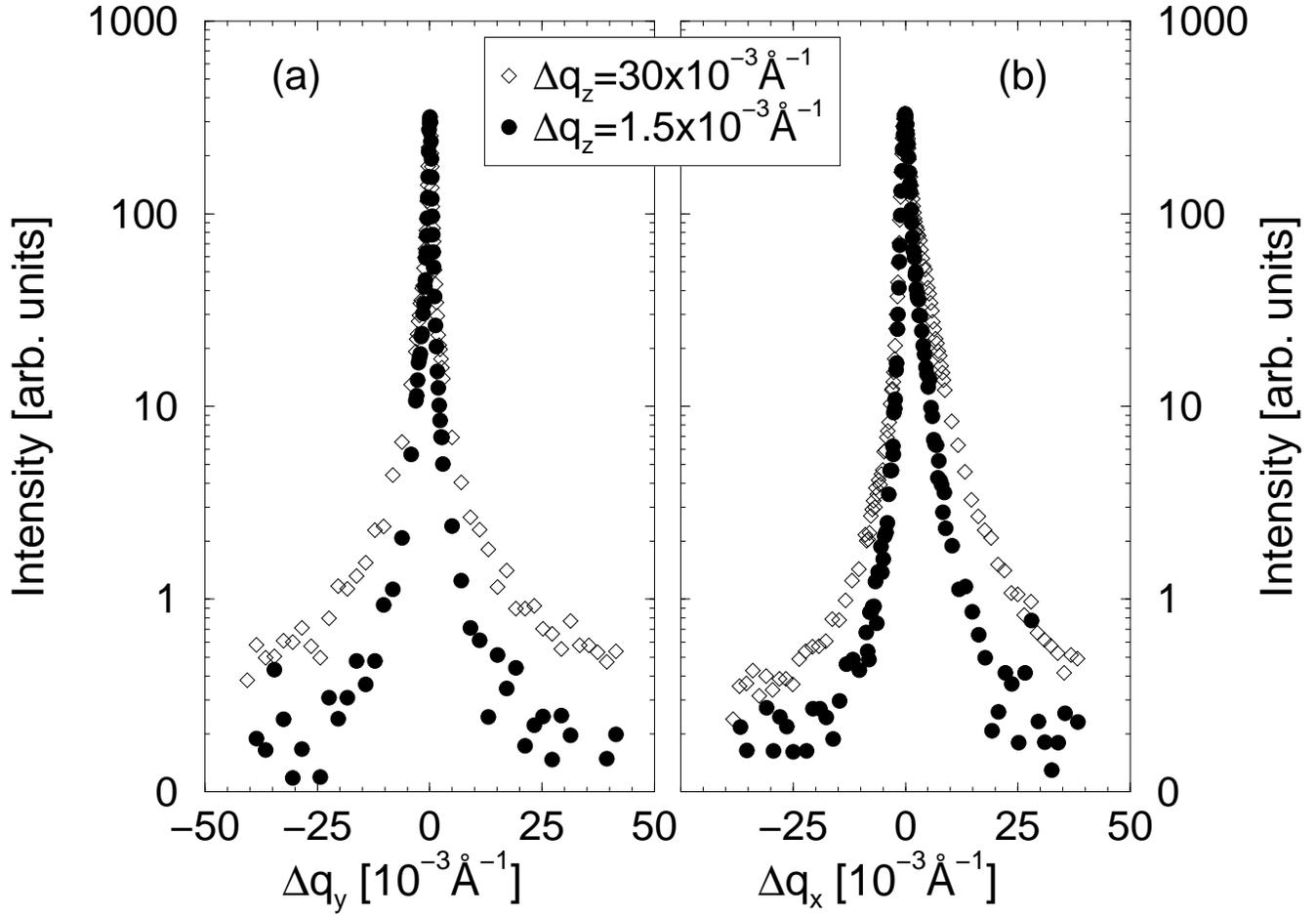}
    \caption{Suppression of the broad component due to improvement of the resolution vertical to the scattering plane shown for transverse (a) and longitudinal (b) scans. Filled circles refer to scans with \Dq{z}$=1.5$\ang{-3}, open diamonds represent scans with \Dq{z}$=30$\ang{-3}. \label{fig:vert}}
\end{figure}

In order to check the validity of the theoretical calcuations, two
different values for the vertical resolution have been set up, without
changing any parameter of the sample and its surrounding.  By
narrowing the detector slit the vertical resolution has been improved
by a factor of $20$.  As shown in figure~\ref{fig:vert}, the
improvement of the resolution suppresses the intensity of the broad
component much more than the intensity of the sharp component in the
center of the scan.  However, applying the fitting procedure described
above the fitted values for the correlation lengths and the
susceptibilities were independent of the resolution width.  This
observation we take as a verification of the calculations presented in
this section.
\newpage
\bibliographystyle{prsty}

\newpage
\newpage
\newpage
\newpage
\newpage
\newpage
\newpage
\newpage
\newpage
\newpage
\newpage
\newpage
\newpage
\newpage
\newpage
\newpage
\begin{table}[!h]
    \centering
    \begin{tabular}[h]{l|l|c|c}
      Sample \#&Growth method&$n_O$ [cm$^{-3}$]&\tc{} [K]\\
      \hline
      \hline
      I&Float-zone grown&$6.1(2)\times 10^{18}$&98.8(2)\\
      \hline
      II&Flux grown&$2.8(2)\times 10^{18}$&102.6(2)\\
      \hline
      III&Verneuil (oxidised)&$7.4(2)\times 10^{16}$&105.7(2)\\
      IV&Verneuil (as grown)&$7.6(2)\times 10^{16}$&105.8(2)\\
      V&Verneuil (reduced)&$1.7(1)\times 10^{19}$&101.0(2)\\
    \end{tabular}
    \caption{Nomenclature of the different samples, concentrations of oxygen vacancies and critical temperatures. \label{tab:samples}}
\end{table}
\newpage
\begin{table}[!h]
    \centering
    \begin{tabular}[h]{l|c|c|c|c}
      &\multicolumn{4}{c}{Broad component}\\
      sample \#&\nb&\gb&\gdnb&$\beta=(3\nu_b-\gamma_{b})/2$\\
      \hline
      \hline
      Original block&&&&\\
      I, region~E& 1.07(6) & 2.9(2)  & {\bf 2.71(24)}&0.16(13)\\
      I, region~A& 1.02(4) & 2.32(7) & {\bf 2.27 (11)}&0.37(7)\\
      \hline
      Residual plate&&&&\\
      I, region~B&1.19&2.23(10)&{\bf 1.87(11)}&0.67(8)\\
      I, region~C&1.19&2.83(10)&{\bf 2.38(13)}&0.37(8)\\
      \hline
      Residual block&&&&\\
      I, region~D&1.19&2.53(3)&{\bf 2.13(9)}&0.52(6)\\
      I, region~E&1.19(4)&2.89(4)&{\bf 2.43(9)}&0.34(6)\\
      \hline
      II&0.9(1)&1.7(1)&{\bf 1.9(2)}&0.50(16)\\
      \hline
      III& 0.73(7) & 1.49(15) & {\bf 2.04(28)}&0.35(13) \\
      IV& 0.79(2) & 1.58(7)  & {\bf 2.00(10)}&0.40(5) \\
      V& 1.18(3) & 2.45(7)  & {\bf 2.08(8)}&0.55(6) \\
    \end{tabular}
    \caption{Critical exponents \nb{} and \gb{} for the broad component. Due to the scaling relations the ratio \gdnb{} should be close to $2.0$. Also, $\beta=(3\nu_b-\gamma_b)/2$ is calculable using the scaling relations. \label{tab:broad}}
\end{table}
\newpage
\begin{table}[!h]
    \centering
    \begin{tabular}[h]{l|c|c|c}
      &\multicolumn{3}{c}{Sharp component}\\
      sample \#&\ns&\gs&\gdns\\
      \hline
      \hline
      I, region~A& 1.17(4)&5.33(9)&{\bf 4.6(2)}\\
      I, region~B&0.58(5)&1.9(2)&{\bf 3.3(4)}\\
      I, region~D&1.30(5)&4.29(10)&{\bf 3.30(14)}\\
      \hline
      II&1.1(1)&4.8(1)&{\bf 4.4(4)}\\
    \end{tabular}
    \caption{Critical exponents \ns{} and \gs{} as well as the ratio \gdns{} for the sharp component. The ratio \gdns{} is more likely close to $4$ than to the expected value of $2$. \label{tab:sharp}}
\end{table}
%
%
\end{document}